\documentclass[a4paper,fleqn]{cas-sc}

\usepackage{silence}
\usepackage[authoryear]{natbib}

\usepackage{subcaption}
\usepackage{graphicx}

\usepackage{threeparttable}
\usepackage{tabularx}
\usepackage{array}
\usepackage{booktabs}
\usepackage{multirow}
\usepackage{makecell}
\usepackage{tcolorbox}
\usepackage[table]{xcolor}
\usepackage{makecell}

\usepackage{algorithmic}
\usepackage[ruled,vlined]{algorithm2e}
\SetKwInput{KwInference}{Inference}
\SetKwInput{KwTraining}{Training}

\usepackage[final]{microtype}

\newcolumntype{Y}{>{\centering\arraybackslash}X}
\newcolumntype{Z}{>{\arraybackslash}X}

\def\tsc#1{\csdef{#1}{\textsc{\lowercase{#1}}\xspace}}
\tsc{WGM}
\tsc{QE}
\tsc{EP}
\tsc{PMS}
\tsc{BEC}
\tsc{DE}

\begin{document}
\let\WriteBookmarks\relax
\def\floatpagepagefraction{1}
\def\textpagefraction{.001}
\shorttitle{Explainable speech deepfake source tracing}
\shortauthors{H. H. Pham et~al.}

\title [mode = title]{Explainability by Design: Structured Kolmogorov–Arnold Networks over Probabilistic Attributes for Speech Deepfake Source Tracing}                      

\tnotetext[1]{This research is the results of the SPEECHFAKES project funded by the Academy of Finland.}

\author[1]{Hoang H. Pham}[orcid=0009-0007-4601-9216]
\cormark[1]
\ead{hpham@uef.fi}
\credit{Writing – review \& editing, Writing – original draft, Validation, Software, Methodology, Investigation, Formal analysis, Data curation, Conceptualization}

\author[1]{Manasi Chhibber}
\ead{manasi.chhibber@uef.fi}
\credit{Writing – review \& editing, Writing – original draft, Methodology, Investigation, Formal analysis}

\author[1]{Tomi H. Kinnunen}
\ead{tomi.kinnunen@uef.fi}
\credit{Writing – review \& editing, Validation, Supervision, Project administration, Methodology, Investigation, Funding acquisition, Formal analysis, Data curation, Conceptualization}

\affiliation[1]{organization={School of Computing, University of Eastern Finland},
                city={Joensuu},
                postcode={FI-80101},
                country={Finland}}

\cortext[cor1]{Corresponding author}

\date{\today}

\begin{abstract}
Modern speech synthesizers can produce highly realistic speech, making source tracing (i.e. identifying the generator behind a spoofed utterance) increasingly important for forensics, online content provenance, and platform accountability. Building on our prior work on transparent probabilistic attributes, which represent utterances as probability distributions over synthesizer sub-components, we extend speech deepfake source tracing with two key ingredients: multi-task training of the probabilistic attribute extractors and a structured Kolmogorov--Arnold Network (KAN) for attack classification. The probabilistic features are estimated jointly with a multi-task learning module built on a shared AASIST or SSL-AASIST countermeasure backbone. The resulting probabilistic feature embedding is classified by a structured KAN whose topology follows known attribute-to-attack relationships. This provides interpretability by construction: the architecture reflects the generative hierarchy of attacks, while KAN feature-importance scores quantify each probabilistic feature's contribution without post-hoc explainers such as SHAP. On ASVspoof2019-attr-17, the extended framework achieves balanced accuracies above $99\%$ for all seven probabilistic feature extractors, with EERs of $0.16\%$--$0.07\%$, and $99.64\%$ balanced accuracy with $0.11\%$ EER for 17-class attack classification. Our revised model outperforms the earlier two-stage baselines, in addition to demonstrating reliable interpretability, with importance scores consistent with SHAP values, and stable results across batch sizes. These findings highlight the potential of structured KAN for speech deepfake source tracing that is both accurate and interpretable by design. For transparency and reproducibility, our codebase is publicly available: \url{https://github.com/HoangHPham/KAN-Probabilistic-Deepfake-Attribution}.
\end{abstract}




\maketitle

\section{Introduction}

Recent advances in generative artificial intelligence (GenAI), including speech synthesis technology \citep{speech_synthesis__ref_1, speech_synthesis__ref_2}, have substantially improved convenience, accessibility, and efficiency in our daily lives. Despite its many beneficial applications, speech synthesis technology has also increased risks related to information security and system reliability, including threats ranging from identity theft and misinformation spread to defamation campaigns and financial fraud \citep{deepfake_risk_ref_1, deepfake_risk_ref_2}. These risks have heightened concerns about voice spoofing attacks, which deliberately aim to deceive human listeners or automatic speaker verification systems. In such exploitative contexts, the contemporary expression \emph{speech deepfake} refers to an artificial or synthesized speech sample generated using a deep learning model.

Detecting the presence of speech deepfakes is crucial for protecting both individuals and organizations against attacks involving the use of deepfakes. 
In response, the research community has devoted substantial effort to developing deepfake detection systems, commonly known as \emph{anti-spoofing} systems or \emph{countermeasures} (CMs). These systems address a binary hypothesis testing problem: given a speech utterance $X$, the objective is to determine whether $X$ is more likely to originate from a bona fide (real) human speech production system (null hypothesis, $H_0$) or from a speech synthesis system (alternative hypothesis, $H_1$).

However, the increasing diversity and sophistication of spoofing attacks demand more reliable and adaptable countermeasures. In this context, \emph{speech deepfake source tracing} 
extends the scope of binary detection by aiming to identify the specific generative model, algorithm(s), or software used to produce a given synthetic recording. Hence, in source tracing one has already established that $X$ does \emph{not} originate from a bona fide human being, with the aim being to identify its source generator. Formally, the source tracing task can be formulated as a multi-class problem: given a spoofed speech utterance $X$ and a set of known attacks (source generators) $\mathscr{A}=\{\text{a}_1, \text{a}_2, \dots, \text{a}_n\}$, the goal is to determine which attack generated $X$, or to infer that it originates from an unknown source not included in $\mathscr{A}$. Accordingly, the source tracing problem can be considered under either \emph{closed-set} or \emph{open-set} conditions. In the former case, $X$ is assumed to originate from one of the known attack methods in $\mathscr{A}$. In contrast, the open-set scenario allows $X$ to be generated by previously unseen attacks outside the set $\mathscr{A}$. We focus on the closed-set formulation.

\begin{figure}
    \centering
    \includegraphics[width=0.75\linewidth]{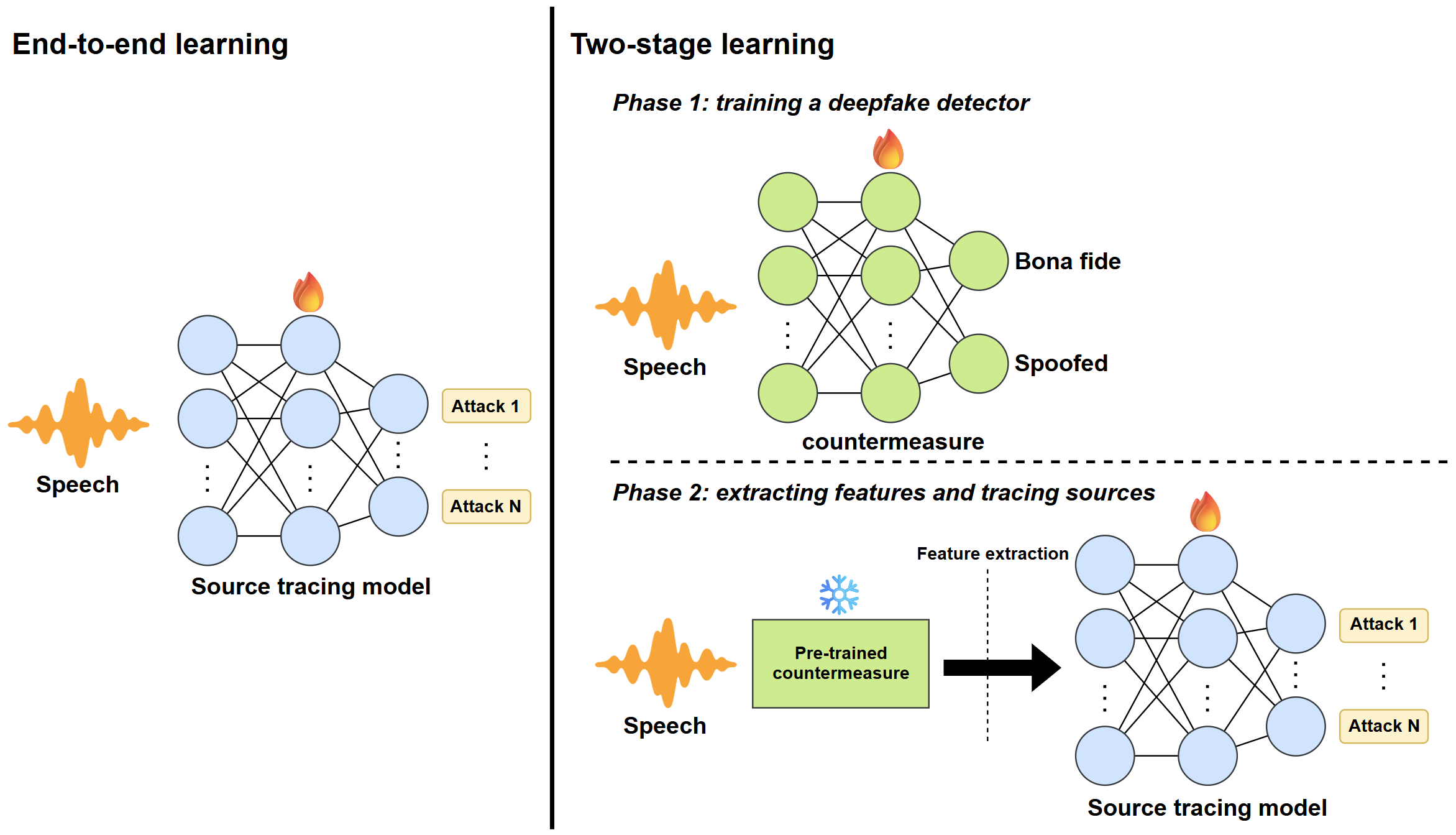}
    \caption{Architectural design strategies for source tracing system (adapted from \citet{klein24_interspeech}).}
    \label{fig:source_tracing_architecture_design}
\end{figure}

From the literature, we identify two primary ways for designing source tracing systems~\citep{klein24_interspeech}, as illustrated in Fig.~\ref{fig:source_tracing_architecture_design}. As in other speech classification tasks, an \emph{end-to-end} design is understood as a monolithic model optimized directly for the given task on one go, here, for identifying the known source generator (attack). In contrast, in the \emph{two-stage} design, a binary system is first trained for the detection task of distinguishing between bona fide and spoofed speech. The representations extracted from this pre-trained model are then used as inputs to a back-end classifier to identify the sources of spoofed samples. In this work, we develop an end-to-end source-tracing model that jointly classifies the attack system and its sub-components, such as waveform generators (vocoders).

The focus of our work is to advance \textbf{explainable} speech deepfake source tracing. Our work builds upon recent efforts to incorporate XAI into speech anti-spoofing, moving beyond simple heatmaps toward structural transparency; a review of related work is provided in Section~\ref{sec:background-XAI}. In particular, our work builds upon (and substantially extends) our recent methodology~\citep{manasi_explainable,mishra_towards} based on \textbf{high-level probabilistic characterization of synthetic speech generators}. The idea is simple: we replace abstract high-dimensional latent space embeddings (whose dimensions are generally meaningless to humans) with \emph{probabilistic features} (that can be readily interpreted). The probabilistic features can then be used in the same way as other features, i.e. combined with a back-end classifier to identify the source generator. Concretely, the probabilistic features used in~\citet{manasi_explainable,mishra_towards} represent the uncertainty of a particular synthesizer sub-component (e.g. a particular waveform generator, duration model, or acoustic feature predictor) being present in the given recording $X$. In terms of operationalization, ~\citet{manasi_explainable,mishra_towards} utilized a bank of independently-trained feedforward networks to estimate the probabilistic features. Following the two-stage design philosophy, the probabilistic feature extractors used the same binary detector embeddings. 

While in our earlier work \citep{manasi_explainable,mishra_towards} we successfully demonstrated that probabilistic attribute embeddings perform comparably to raw latent-space features, the underlying architecture was inherently fragmented. Specifically, these frameworks relied on a bank of independently trained multi-layer perceptrons (MLPs) for attribute extraction, followed by separate back-end classifiers (e.g., decision trees \citep{decision_tree} or logistic regression \citep{logistic_regression}). Furthermore, they depended on post-hoc Shapley value (SHAP) \citep{Shapley} estimations to quantify feature importance. This disjointed pipeline presents fundamental limitations: not only do basic classifiers like decision trees struggle with stability and large-scale accuracy \citep{decisiontree_limit}, but relying on external post-hoc explainers increases computational overhead and obscures whether the generated explanations faithfully reflect the model's true internal logic. Consequently, while we established a necessary conceptual foundation, there remains a critical need for a unified, high-performance architecture where explainability is an intrinsic structural feature rather than an afterthought.

To bridge this gap, the present work proposes a novel, end-to-end architecture that jointly optimizes feature representation, attribute extraction, and attack classification. Moving away from independently trained MLPs, we integrate \emph{multi-task learning} (MTL) with a specialized \emph{Kolmogorov–Arnold network} (KAN). In this unified framework, the MTL module leverages a countermeasure backbone to learn shared, globally informed speech representations that are used to train probabilistic attribute extractors. Probabilistic attribute embeddings are then directly feed into our proposed \emph{structured KAN module} (SKM), bypassing the need for black-box back-ends. The adoption of KAN serves as our primary structural contribution, yielding both intrinsic and extrinsic interpretability by design. Specifically, the SKM explicitly wires the relationships between attacks and their corresponding generative attributes (derived from ASVspoof 2019 LA metadata) into the network's topology, ensuring complete structural transparency. Furthermore, KAN features a built-in importance analysis mechanism that directly quantifies the contribution of each attack attribute without relying on external post-hoc tools. Alongside standard classification metrics, the robustness of our integrated interpretability mechanism is further validated through targeted consistency and stability evaluations.

\begin{figure}
    \centering
    \includegraphics[width=1\linewidth]{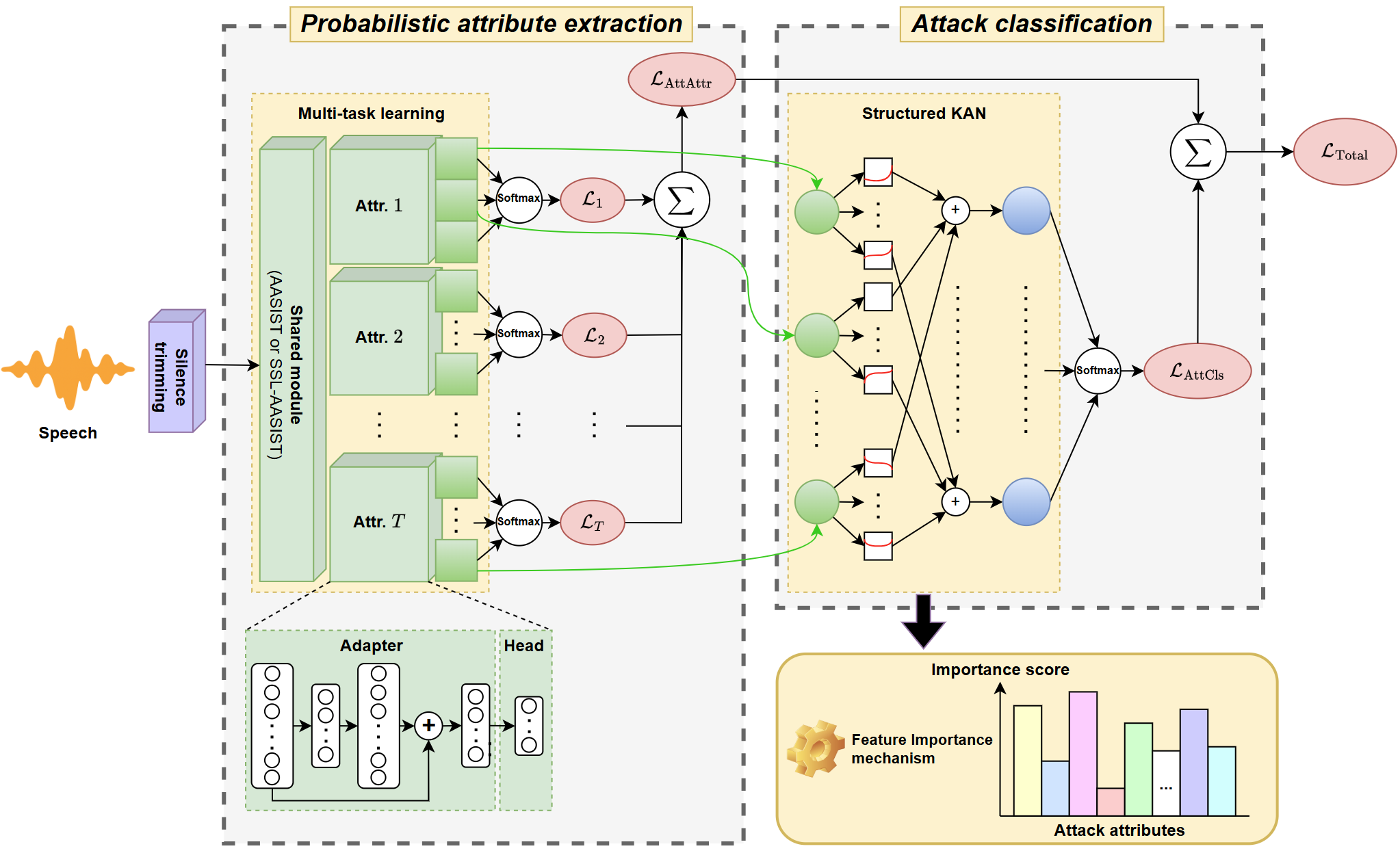}
    \caption{Proposed end-to-end source tracing model for speech deepfakes in the ASVspoof 2019 LA dataset. The multi-task learning (MTL) module, consisting of a shared backbone and $T$ attribute extractors (Attr. $i$), predicts attack attributes, while the structured Kolmogorov–Arnold network module (SKM) performs attack classification. A built-in interpretability mechanism is further employed to assess the contribution of each attribute in the classification process.}
    \label{fig:proposed_model}
\end{figure}

\section{Background: Explainability}\label{sec:background-XAI}

\subsection{Explainable Artificial Intelligence (XAI)}

 
Whereas advances in deep learning have paved way for many new applications---including speech deepfake detection and source tracing---it is also widely acknowledged that pure black-box decisions are insufficient in many cases. In critical domains such as forensics or medicine, the reasoning and transparency of decision making are vital. In this context, the umbrella term \emph{explainable artificial intelligence} (XAI)~\citep{XAI_ref} refers to approaches for "opening the black box", typically in the context of deep neural network models. Even if the two terms are often used exchangeably, in XAI one distinguishes between interpretability and explainability. \textbf{Interpretability} refers to the extent to which the internal structure of a model is understandable, while \textbf{explainability} refers to the ability to provide reasons for model decisions. This leads to two distinguishable terms: (1) \emph{intrinsic} and (2) \emph{extrinsic} interpretability \citep{interpretability}. In the former, the model structure itself provides insight into its internal functioning, whereas the latter relates to the rationale behind the model decisions through post-hoc analyses, such as input feature importance estimation or heatmap-based visualization techniques.

In practice, extrinsic interpretability is often achieved using post-hoc explanation methods that analyze a trained model after training. These methods estimate the contribution of input to the prediction of a model. Common approaches include \emph{integrated gradients} (IG) \citep{XAi_integrated_gradient}, which attributes importance by accumulating gradients from a baseline input; \emph{local interpretable model-agnostic explanations} (LIME) \citep{XAI_LIME}, which approximates the local behavior of a complex model using a simpler surrogate model; and \emph{shapley additive explanations} (SHAP) \citep{Shapley}, which assigns feature importance values based on a game-theoretic formulation. These techniques do not change the model itself but provide post-hoc insights into its decisions. In contrast, intrinsic interpretability is achieved by designing models that are transparent by structure, such as classical interpretable models like decision trees or random forests, as well as more recent explainable neural architectures, including the \emph{speech formant audio transformer network} (SFATNet) \citep{SFATNet_1, SFATNet_2}, which is designed to be interpretable by design. SFATNet uses frame-level segmentation along the time axis, where each token represents a full frequency frame, enabling frame-level analysis. It also applies multi-head attention pooling to assign importance weights to different time frames, indicating their contribution to the final decision. In this work, we address both intrinsic and extrinsic interpretability within a unified framework.

%

\subsection{Explainable AI (XAI) for speech anti-spoofing}

\begin{table*}
\centering
\caption{Summary of related works on interpretable speech deepfake detection and source tracing. Abbreviations: SHAP = SHapley Additive exPlanations, LIME = Local Interpretable Model-agnostic Explanations, Grad-CAM = Gradient-weighted Class Activation Mapping, SFATNet = Speech Formant Audio Transformer Network and KAN = Kolmogorov–Arnold Network.}
\label{tab:related_work_xai_in_speech}
\renewcommand{\arraystretch}{1.2}
\setlength{\tabcolsep}{5pt}
\small

\begin{threeparttable}
\begin{tabularx}{\textwidth}{l l l l X}
\toprule
\textbf{Paper} & \textbf{Task} & \textbf{Dataset} & \textbf{Interpretability} & \textbf{XAI Method} \\ \hline

\midrule
\citet{Related_work_XAI_1} & Detection & \makecell[l]{- ASVspoof 2021 LA \\ - LJSpeech} & Extrinsic & \makecell[l]{- Integrated gradients \\ - Taylor decomposition} \\

\midrule
\citet{Related_work_XAI_4} & Detection & \makecell[l]{ASVspoof 2019 LA} & Extrinsic & SHAP \\

\midrule
\citet{Related_work_XAI_2} & Detection & \makecell[l]{ASVspoof 2019 LA} & Extrinsic & SHAP \\

\midrule
\citet{Related_work_XAI_3} & Detection & \makecell[l]{- ASVspoof 2019 LA \\ - LJSpeech} & Extrinsic & \makecell[l]{- Layer-wise relevance propagation \\ - Integrated gradients \\ - Deep Taylor decomposition} \\

\midrule
\citet{Related_work_XAI_5} & Detection & \makecell[l]{- ASVspoof 5 \\ - In-the-Wild \\ - FakeOrReal \\ - TIMIT-TTS} & Intrinsic & SFATNet-based design \citep{SFATNet_1, SFATNet_2} \\

\midrule
\citet{Related_work_XAI_6} & Detection & \makecell[l]{- ASVspoof 2019 LA \\ - ASVspoof 2019 PA \\ - In-the-Wild \\ - FakeAVCelebV2} & \makecell[l]{Intrinsic \& \\ Extrinsic} & \makecell[l]{- ML-based designs (KNN, random \\ forest, XG Boost) \\ - SHAP} \\

\midrule
\citet{Related_work_XAI_7} & Detection & \makecell[l]{- ASVspoof 5 \\ - FakeAVCeleb} & \makecell[l]{Intrinsic \& \\ Extrinsic} & \makecell[l]{- Gradient boosting decision tree \\ based feature importance analysis\\ - Attention visualization} \\

\midrule
\citet{Related_work_XAI_8} & Detection & ASVspoof 2021 LA & Extrinsic & \makecell[l]{- SHAP \\ - LIME \\ - Grad-CAM} \\

\midrule
\citet{Related_work_XAI_9} & Detection & In-the-Wild & Extrinsic & LIME \\

\midrule
\citet{xixuan_wstx} & Detection & \makecell[l]{- Deepfake-Eval-2024 \\ - SpoofCeleb \\ - In-the-Wild} & Extrinsic & SHAP \\

\midrule
\citet{Related_work_XAI_10} & Source tracing & CodecFake+ & Extrinsic & Attention visualization \\

\midrule
\citet{manasi_explainable} & \makecell[l]{Detection \& \\ Source tracing} & \makecell[l]{ASVspoof 2019 LA \\ (2 attacks)} & \makecell[l]{Intrinsic \& \\ Extrinsic} & \makecell[l]{- Decision tree \\ - SHAP} \\

\midrule
\citet{mishra_towards} & \makecell[l]{Detection \& \\ Source tracing} & \makecell[l]{ASVspoof 2019 LA \\ (17 attacks)} & \makecell[l]{Intrinsic \& \\ Extrinsic} & \makecell[l]{- Decision tree \\ - SHAP} \\

\midrule
\rowcolor{green!10}
\textbf{This work} & \makecell[l]{Detection \& \\ Source tracing} & \makecell[l]{ASVspoof 2019 LA \\ (17 attacks)} & \makecell[l]{Intrinsic \& \\ Extrinsic} & \makecell[l]{- Structured KAN \\ - KAN's built-in mechanism for \\ feature importance analysis} \\

\bottomrule
\end{tabularx}
\end{threeparttable}

\end{table*}

Anti-spoofing in speech technologies has become a significant research area, with recent advances focusing on improving interpretability of system. Table \ref{tab:related_work_xai_in_speech} summarizes prior works on interpretability for both detection and source-tracing tasks.

Many studies applied post-hoc explanation methods to analyze the outputs of deepfake detectors. For example, \citet{Related_work_XAI_1} and \citet{Related_work_XAI_3} used gradient-based methods such as integrated gradients \citep{XAi_integrated_gradient}, layer-wise relevance propagation \citep{XAI_layer_wise_relevance_propagation}, and Taylor decomposition \citep{XAI_Taylor_decomposition}. Other works estimated feature importance using SHAP or LIME \citep{Related_work_XAI_4, Related_work_XAI_2, Related_work_XAI_8, Related_work_XAI_9, xixuan_wstx}. Some studies instead focus on intrinsic interpretability through model design or interpretable machine learning architectures, including the SFATNet-based model \citep{Related_work_XAI_5} and classical models such as decision tree or random forest combined with explanation tools \citep{Related_work_XAI_6, Related_work_XAI_7}.

Beyond detection, far fewer studies examine interpretability for the source tracing task. Examples include the attention-based method \citep{Related_work_XAI_10} and the use of a decision tree with SHAP analysis in \citet{manasi_explainable} and \citet{mishra_towards}. Despite these efforts, research that jointly ensures \emph{both} intrinsic \emph{and} extrinsic interpretability in speech deepfake systems remains limited, particularly in source tracing. In addition, post-hoc methods increase computational cost, while simple interpretable models often underperform. We therefore design a source-tracing system with both intrinsic and extrinsic interpretability, without relying on any post-hoc method.

\subsection{Kolmogorov-Arnold Network for speech anti-spoofing}

\begin{table*}
\centering
\caption{Summary of related works on the use of KAN in speech anti-spoofing.}
\label{tab:related_work_kan_speech_anti_spoofing}
\renewcommand{\arraystretch}{1.2}
\setlength{\tabcolsep}{5pt}
\small

\begin{threeparttable}
\begin{tabularx}{\textwidth}{l l l l X}
\toprule
\textbf{Paper} & \textbf{Task} & \textbf{Dataset} & \textbf{Interpretability} & \textbf{Role of KAN} \\
\hline
\midrule
\citet{Related_work_KAN_antispoofing_1} & Deepfake detection & \makecell[l]{- ASVspoof 2019 LA \\ - ASVspoof 2021 LA, DA} & No & Feature projector \\

\midrule
\citet{Related_work_KAN_antispoofing_2} & Deepfake detection & \makecell[l]{- ASVspoof 2019 LA \\ - ASVspoof 2021 LA, DA \\ - In-the-Wild} & No & Feature projector \\

\midrule
\citet{Related_work_KAN_antispoofing_3} & Deepfake detection & ASVspoof 5 & No & Feature projector \\

\midrule
\citet{Related_work_KAN_antispoofing_4} & Deepfake detection & ASVspoof 2021 LA & No & Feature projector \\

\midrule
\citet{Related_work_KAN_antispoofing_5} & Voice liveness detection & \makecell[l]{- POCO \\ - ASVspoof 2017 V2 \\ - ASVspoof 2019} & No & Classifier \\

\midrule
\rowcolor{green!10} \textbf{This work} & Deepfake source tracing & \makecell[l]{ASVspoof 2019 LA \\ (17 attacks)} & Yes & Classifier \\

\bottomrule
\end{tabularx}
\end{threeparttable}

\end{table*}

\emph{Kolmogorov-Arnold network} (KAN) introduced recently by~\cite{KAN1.0} is a neural network architecture that provides an alternative to conventional multilayer perceptrons. Before providing a self-contained introduction to KAN architecture (Section~\ref{section:preliminary_KAN}), we provide a brief review of its typical usage, including prior studies in speech anti-spoofing. 
In practice, KAN has been integrated into several deep learning architectures and has been applied in many domains beyond speech, including computer vision, time series analysis, and graph learning. In transformer-based models introduced by \citet{kan_transformer_1} and \citet{flash_kan_transformer}, KAN layers are typically used to replace MLP blocks in feed-forward networks to improve nonlinear transformation. Similarly, in CNN and ResNet-based architectures, KAN is applied as a substitute for fully connected layers or convolutional components, leading to variants such as Convolutional-KAN \citep{cnn_kan_1} and Residual-KAN \citep{residual_kan}.

In speech anti-spoofing, the use of KAN remains limited, as seen from the summary in Table~\ref{tab:related_work_kan_speech_anti_spoofing}. Prior work has mainly explored two configurations: using KAN as a feature projector to learn intermediate representations, or applying it only in the final classification layer \citep{kan_speech_anti_spoofing}. For example, \citet{Related_work_KAN_antispoofing_3} integrate KAN into the well-known AASIST deepfake detection model introduced by ~\cite{Jung2021AASIST}. However, these studies treat KAN largely as a drop-in replacement for MLP, leaving the overall network architecture largely unchanged and focusing primarily on detection performance rather than interpretability. More broadly, existing work typically replaces the linear transformations and fixed activation functions of MLP with KAN’s learnable activations while retaining the surrounding architecture, such as a transformer, CNN, ResNet, or AASIST backbone. Consequently, KAN has so far been evaluated mainly for predictive performance. In contrast, \citet{Liu2024KAN2K} introduce a built-in mechanism to directly estimate input feature importance from the model. This provides an alternative to post-hoc interpretability methods such as SHAP and helps address an important gap in speech anti-spoofing research.

\section{Methodological Preliminaries}

\subsection{Kolmogorov-Arnold Network (KAN)} \label{section:preliminary_KAN}

While \emph{multi-layer perceptron} (MLP) \citep{MLPs} are grounded in the \emph{universal approximation theorem} (UAT) \citep{UAT}, \emph{Kolmogorov-Arnold network} (KAN) \citep{KAN1.0} rely on the \emph{Kolmogorov-Arnold representation theorem} (KART) \citep{KART} to approximate functions of multivariate inputs. Specifically, UAT guarantees approximation of a target function through compositions of \emph{multivariate} functions with suitable parameters, whereas KART guarantees approximation through compositions of continuous \emph{univariate} functions. The key difference between the computational schemes in Equations \eqref{eq_uat_one_hidden_nn} and \eqref{eq_orig_kart} is therefore how they represent nonlinearity.

\begin{tcolorbox}[colback=green!5!white,colframe=black!75!black,title=Key difference of representation theorems relevant to MLP and KAN.]
\emph{Theorem 3.1 (Universal approximation theorem, UAT).} Let $\mathbf{x} \in \mathbb{R}^{m}$ be the input vector, let $\sigma$ be a fixed nonlinear activation function (e.g., sigmoid, tanh, or ReLU), and let $\mathbf{w} \in \mathbb{R}^{m}$ and $b \in \mathbb{R}$ denote the learnable weight vector and bias, respectively. Then the finite sum
\begin{equation}\label{eq_uat_one_hidden_nn}
    f_{\mathrm{NN}}(\mathbf{x}) = \sum_{i=0}^{N-1} \sigma(\mathbf{w}_{i}^\intercal \mathbf{x} + b_{i}) ~,
\end{equation}
defines the approximation of a function $f$ by a one-hidden-layer neural network with $N$ hidden neurons.
\newline

\emph{Theorem 3.2 (Representation theorem, KART).} Let $f \colon [0,1]^m \rightarrow \mathbb{R}$ be a continuous multivariate function, and let $\mathbf{x} \in [0,1]^m$ be the input vector. Then $f$ can be represented as
\begin{equation}\label{eq_orig_kart}
f(\mathbf{x}) = f(x_{0}, x_{1}, \ldots, x_{m-1}) = \sum_{q=0}^{2m} \Phi_{q}\!\left( \sum_{p=0}^{m-1} \phi_{q,p}\!\left(x_{p}\right) \right) ~,
\end{equation}
where $\Phi_{q} \colon [0,1] \rightarrow \mathbb{R}$ and $\phi_{q,p} \colon \mathbb{R} \rightarrow \mathbb{R}$ are continuous univariate functions.

\end{tcolorbox}

\begin{figure}
    \centering
    \includegraphics[width=0.75\linewidth]{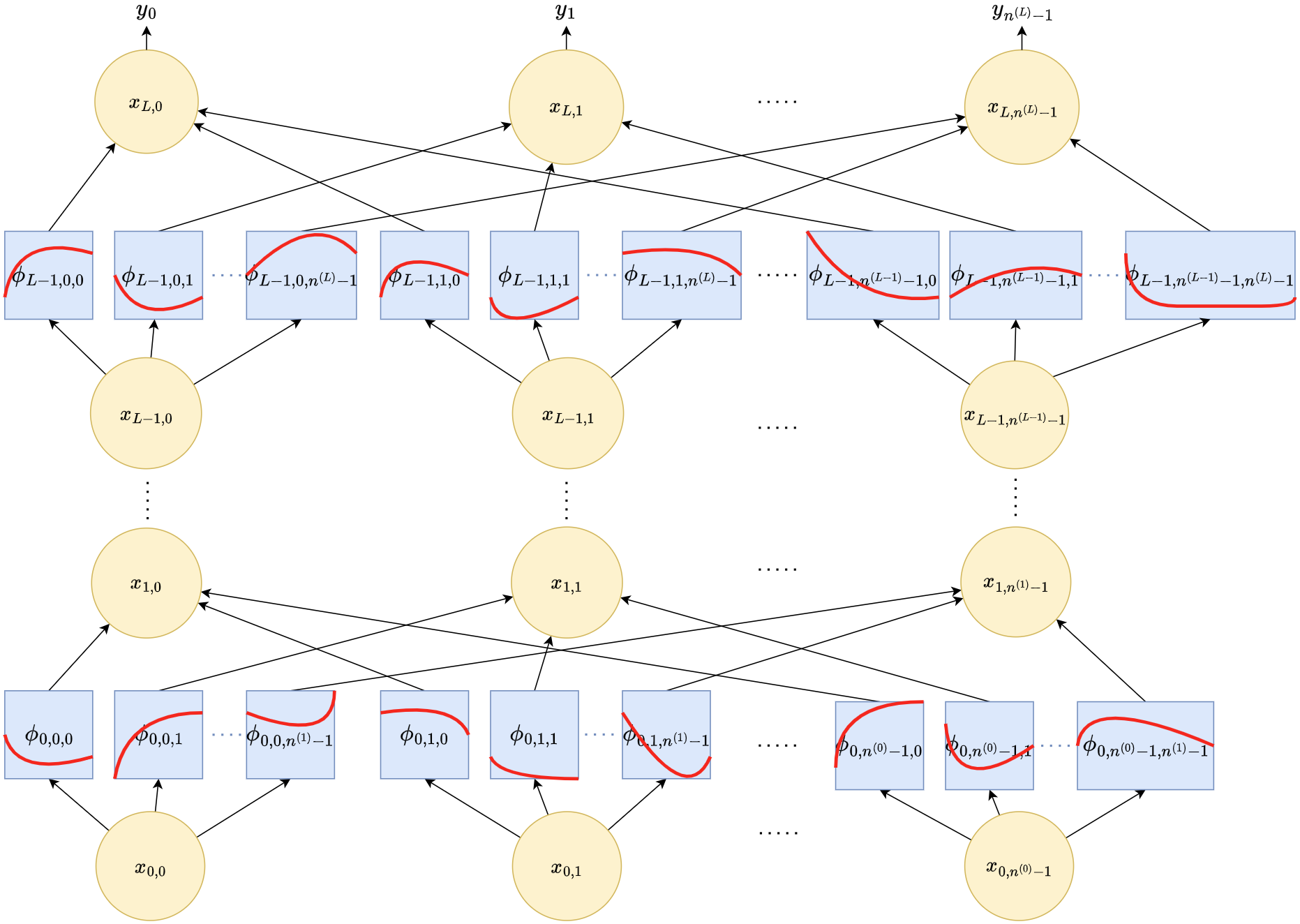}
    \caption{Computational graph of multilayer KAN model.}
    \label{fig:kan_overall}
\end{figure}

The representation theorem in Eq. \eqref{eq_orig_kart} shows that \((2m+1) \times m\) inner univariate functions \(\phi_{q,p}\) and \(2m+1\) outer univariate functions \(\Phi_{q}\) are sufficient to represent exactly the given \(m\)-variate function. When considering KART in the context of a neural network, the Equation \eqref{eq_orig_kart} follows precisely the form of a function approximation performed by a one-hidden-layer neural network. However, similar to MLP, this shallow form limits the expressive power of the network. This has motivated generalizations of Equation~\eqref{eq_orig_kart} to wider and deeper architectures, leading to the development of KAN. The key idea inherited from KART is the use of \emph{learnable} univariate functions $\phi$. Instead of applying a linear transformation followed by a fixed nonlinear activation at each node (as in MLP), KAN places \emph{parameterized} univariate activation functions on the edges. In the original KAN study by~\citet{KAN1.0}, each univariate function is parameterized by a \emph{B-spline curve} \citep{chaudhuri2021bsplines}, whose local B-spline basis coefficients are learned during training. Figure~\ref{fig:kan_overall} illustrates the overall architecture of a multilayer KAN model.

The structure of an $L$-layer KAN is specified by an array $[n^{(0)}, n^{(1)}, \ldots, n^{(L)}]$, where $n^{(0)}$ is the input dimension, $n^{(L)}$ is the output dimension, and $n^{(l)}$ for $l=1,\ldots,L-1$ is the number of nodes in the $l^\mathrm{th}$ hidden layer. Connecting layer $l$ to layer $(l+1)$ requires $n^{(l)} \times n^{(l+1)}$ learnable activation functions, each linking node $i$ in layer $l$ to node $j$ in layer $(l+1)$. We denote each such function by
\begin{equation}\label{eq_kan_activation_function_denotion}
\phi_{l,i,j}; \quad l=0,\ldots,L-1, \quad i=0,\ldots,n^{(l)}-1, \quad j=0,\ldots,n^{(l+1)}-1.
\end{equation}
\noindent Given the univariate input $x_{l,i}$ to $\phi_{l,i,j}$, the corresponding output (post-activation) $\tilde{x}_{l,j,i} \equiv \phi_{l,i,j}(x_{l,i})$ is defined as
\begin{equation}\label{eq_kan_activation_function}
\tilde{x}_{l,j,i} \equiv \phi_{l,i,j}(x_{l,i}) = w_{b_{l,i,j}} b(x_{l,i}) + w_{s_{l,i,j}} \mathrm{spline}_{l,i,j}(x_{l,i}),
\end{equation}
where $b(x_{l,i})$ is the \emph{sigmoid linear unit} (SiLU) function \citep{SiLU}, used as a residual connection:
\begin{equation}\label{eq_kan_basis_function_b(x)}
b(x_{l,i}) = \mathrm{SiLU}(x_{l,i}) = x_{l,i} \cdot \frac{1}{1+e^{-x_{l,i}}} ~,
\end{equation}
and where $\mathrm{spline}_{l,i,j}(x_{l,i})$ denotes a $B$-spline defined by:
\begin{equation}\label{eq_kan_spline_function}
\mathrm{spline}_{l,i,j}(x_{l,i}) = \sum_{h=0}^{G+k-1} c_h B_h(x_{l,i}).
\end{equation}
Here, $c_h$ are learnable coefficients and $B_h$ are fixed $B$-spline basis functions determined by the spline order and knot configuration \citep{chaudhuri2021bsplines}.

\begin{figure}
    \centering
    \includegraphics[width=0.5\linewidth]{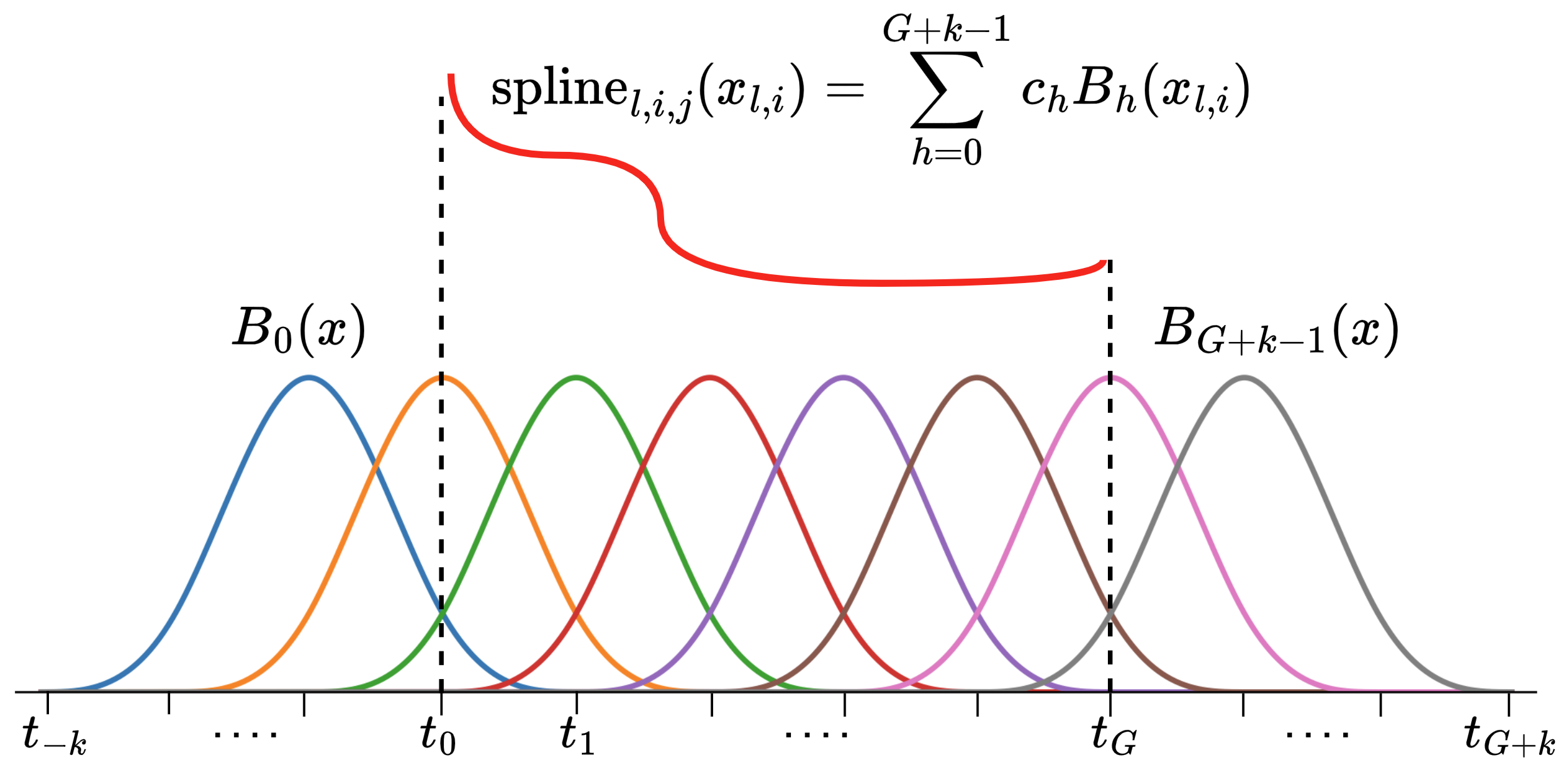}
    \caption{Each learnable univariate activation function in KAN is parameterized from \(k\)-degree B-spline curve defined by \(G\) intervals and weighted sum of basis functions with learnable weights.}
    \label{fig:kan_bspline}
\end{figure}

Figure \ref{fig:kan_bspline} illustrates how a spline in KAN is constructed from the weighted sum of the B-spline basis functions. 
The \(k^{\mathrm{th}}\) degree B-spline is defined by a grid of the \(G\) intervals corresponding to \(G+1\) grid points, also known as \emph{internal knots} \citep{chaudhuri2021bsplines}. At the training time, each learnable activation function \(\phi_{l,i,j}\) is initialized with \(w_{s_{l,i,j}}=1\), \(\mathrm{spline}_{l,i,j}(x_{l,i})\approx 0\) and Xavier initialization \citep{xavier_init} based \(w_{b_{l,i,j}}\). Then, the activation value of the \(j^\mathrm{th}\) node at layer \(l+1\) is calculated by the sum of all incoming post-activations:
\begin{equation}\label{eq_kan_activation_value}
x_{l+1,j} = \sum_{i=0}^{n^{(l)}-1} \tilde{x}_{l,i,j} = \sum_{i=0}^{n^{(l)}-1} \phi_{l,i,j}(x_{l,i}), \quad j=0,...,n^{(l+1)}-1.
\end{equation}
We can present~\eqref{eq_kan_activation_value} more compactly in matrix notation. The activation vector \(\textbf{x}_{l+1} \in \mathbb{R}^{n^{(l+1)}\times1}\) in layer \(l+1\) is a dot product of a matrix of \(n^{(l+1)}\times n^{(l)}\) 1D functions \(\mathbf{\Phi}^{(l)}\) with a pre-activation vector \(\textbf{x}_l \in \mathbb{R}^{n^{(l)}\times1}\) in layer \(l\):
\begin{equation}\label{eq_kan_matrix_form}
\textbf{x}_{l+1}
=
\underbrace{\left(
\begin{array}{cccc}
\phi_{l,0,0}(\cdot) & \phi_{l,1,0}(\cdot) & \cdots & \phi_{l,n_l-1,0}(\cdot)\\
\phi_{l,0,1}(\cdot) & \phi_{l,1,1}(\cdot) & \cdots & \phi_{l,n_l-1,1}(\cdot)\\
\vdots              & \vdots              & \ddots & \vdots              \\
\phi_{l,0,n_{l+1}-1}(\cdot) & \phi_{l,1,n_{l+1}-1}(\cdot) & \cdots & \phi_{l,n_l-1,n_{l+1}-1}(\cdot)
\end{array}
\right)}_{=: \mathbf{\Phi}^{(l)}}
\textbf{x}_{l}.
\end{equation}
Then, given an input vector \(\textbf{x}\), the output of a \(L\)-layer KAN model is calculated by a composition of \(L\) function matrices:
\begin{equation}\label{eq_kan_general_deep_model}
\mathrm{KAN}(\textbf{x}) = (\mathbf{\Phi}^{(l-1)} \circ \mathbf{\Phi}^{(l-2)} \circ \cdots \circ \mathbf{\Phi}^{(1)} \circ \mathbf{\Phi}^{(0)})\textbf{x} ~.
\end{equation}

\subsection{KAN vs MLP: an interpretable alternative}

As noted above, most previous studies treat KAN simply as a replacement for MLP for representation learning. In practice, however, the practical performance difference between KAN and MLP remains unclear. On the one hand, KAN is used due to their ability to potentially achieve faster convergence by replacing linear weight matrices with spline-based functions, which can reduce the number of parameters. On the other hand, several challenges associated with KAN have also been reported, including scalability issues and slow training \citep{kan_transformer_1}. A systematic study conducted by~\cite{KAN_vs_MLP} indicates that, under the same number of parameters, KAN may outperform MLP in symbolic regression tasks but remain inferior in other tasks such as computer vision, natural language processing, and speech processing. 

Our motivation for using KAN lies in its interpretability, rather than predictive performance. In fact, in its first version \citep{KAN1.0}, KAN was introduced as an interpretable architecture due to its learnable B-spline-based activation functions. In KAN 2.0 \citep{Liu2024KAN2K}, a built-in mechanism was further proposed for estimating the contribution of these functions, enabling direct assessment of input feature importance. We focus on leveraging this capability, as opposed to suggesting KAN to replace MLP in terms of predictive performance. We expect that integrating KAN can provide a new direction for explainable AI, which is currently dominated by post-hoc methods.  

\section{Proposed source tracing model}

We propose an end-to-end speech deepfake source tracing architecture that takes spoofed speech as input, extracts attribute values and classifies the corresponding attack simulatenously, as illustrated in Figure \ref{fig:proposed_model}. The proposed architecture comprises three main components: (1) a \emph{multi-task learning module} (MTL), (2) a \emph{structured KAN module} (SKM), and (3) a \emph{built-in feature importance mechanism}. The MTL module extracts probabilistic features, extending our recent works~\citep{manasi_explainable,mishra_towards}. The SKM module, based on a KAN architecture, then serves as a back-end classifier that predicts the source generator label (attack ID). Finally, the feature importance mechanism enabled by the KAN classifier quantifies the relative contribution of the probabilistic features to the attack-label prediction.

\subsection{Multi-task learning (MTL) module: Enhancing Posterior Feature Extraction} \label{section:proposed_MTL}

As background, we represent a speech utterance $X$ through a set of attribute-wise posterior distributions rather than a single opaque embedding. Let $\textbf{h}(X) \in \mathbb{R}^{D}$ denote an utterance-level representation extracted from $X$ by a suitable front-end speech model. Assume that we define $T$ discrete attributes relevant to synthetic speech generation, and that attribute $t$ has $C_{t}$ possible values. For each attribute, we design a feedforward neural network $g_t: \mathbb{R}^D \rightarrow \mathbb{P}^{C_t}$ to predict a vector of attack attribute posteriors\footnote{Here, $\mathbb{P}^{C_t}$ denotes a probability simplex.}:
\begin{equation}\label{eq:posterior-vector}
\mathbf{p}^{(t)}(x) = g_{t}(\boldsymbol{h}(x)) = \bigl[p^{(t)}_{1}(x), \ldots, p^{(t)}_{C_{t}}(x)\bigr]^\intercal
\end{equation}
whose elements are non-negative and sum up to 1. Whereas $t$ indices semantic \emph{attributes} (e.g. vocoder type or duration model architecture) relevant for high-level characterization of the synthesizer, the elements of the posterior vector in~\eqref{eq:posterior-vector} correspond to uncertainty in the predicted \emph{values} of these attributes---a discrete probability distribution over the allowed outcomes (e.g. list of vocoders available at the training time). Stacking all attribute-wise posteriors yields a \emph{probabilistic attribute embedding}. This construction turns the utterance into a compact, semantically structured representation in which each block corresponds to one interpretable attribute and its uncertainty over possible values. 

In our earlier works \citep{manasi_explainable,mishra_towards}, the attribute posterior predictors were realized as a bank of separately trained MLPs on top of spoofing countermeasure embeddings. In contrast, the MTL module introduced in the present work trains the probabilistic attribute extractors \emph{jointly}. The MTL module consists of a shared front-end module along with $T$ attribute-specific blocks, one per attribute. For the former, we adopt two modern deepfake detection architectures, AASIST \citep{Jung2021AASIST} and SSL-AASIST \citep{tak2022automatic}. These are used as backbones where the final linear layer (originally designed for binary classification) is removed. The output of the last hidden layer is used as the shared representation that feeds the attribute specific models.

Concretely, each attribute-specific branch consists of two components: an \emph{attribute adapter} and an \emph{attribute head}. The attribute adapter follows the \emph{Houlsby architecture} \citep{houlsby2019parameter} to learn attribute-specific representations. In our implementation, each adapter applies layer normalization to stabilize shared representations, followed by a bottleneck structure consisting of a down-projection and an up-projection layer that compresses features into a lower-dimensional space and then restores them. GELU activation \citep{Gelu} and dropout are used to introduce nonlinearity and improve generalization. A residual connection is added between the input and transformed features to facilitate stable optimization and preserve shared information. Finally, an output projection layer maps the adapted representations to the attribute-specific feature space. On top of each attribute adapter, an attribute head is implemented as a single linear layer whose output dimension matches the number of values of the corresponding attribute. A softmax function is then applied to produce a probability distribution over the possible attribute values. The architectural details are presented in Section \ref{section:exp_setup_model_configs}. At the training time, each attribute-specific block is optimized using a categorical cross-entropy loss:
\begin{equation}\label{eq_cce_loss}
\mathcal{L}_{t} = - \sum_{i=1}^{C_t}y_{t,i}\log(\hat{y}_{t,i}),
\end{equation}
where \(y_{t,i}\)
is the ground-truth of the value \(i\) of attribute \(t\). The overall (multi-task) loss for probabilistic attribute extraction, denoted by $\mathcal{L}_{\mathrm{AttAttr}}$, is the sum of the losses across all the $T$ attributes, i.e. $\mathcal{L}_{\mathrm{AttAttr}}=\sum_{t=1}^T \mathcal{L}_t$.

\subsection{Structured Kolmogorov-Arnold network module (SKM)}

The second module of the proposed model is a classifier that predicts the attack ID from the probabilistic attribute embedding produced by the MTL module. We adopt a KAN-based classifier for this stage because its learnable univariate functions offer a more transparent mapping from attribute-level evidence to the final class decision, as motivated in Section~\ref{section:preliminary_KAN}. This way, the classifier is intended not only to achieve accurate attack prediction but also to make the relationship between attack attributes and attack classes transparent.

Our proposed \textbf{structured KAN module} (SKM) is implemented as a shallow KAN that maps the probabilistic attribute embeddings into attack posteriors\footnote{technically, into their logits.}. Rather than relying on an unconstrained fully-connected KAN design, our SKM is structured according to prior knowledge about the relationship between attack classes and attribute groups available at the training time (elaborated further in the experimental part). This architectural constraint encourages the model to reflect meaningful dependencies between extracted attributes and predicted attacks, thereby improving interpretability at the model level. The output logits are converted to class probabilities using a softmax function, and the predicted attack is given by the class with the highest posterior probability. The attack classification branch is trained with a categorical cross-entropy loss, denoted by $\mathcal{L}_{\mathrm{AttCls}}$. The overall training objective combines the attribute-extraction loss of the MTL module with the attack-classification loss of the SKM:
\begin{equation}\label{eq_overall_loss}
\mathcal{L}_{\mathrm{Total}} = \mathcal{L}_{\mathrm{AttAttr}} + \mathcal{L}_{\mathrm{AttCls}} ~.
\end{equation}

\subsection{Extrinsic interpretability: a built-in feature importance mechanism}\label{section:proposed_fi_mechanism}

To complement the interpretability of the probabilistic attribute representation, we also exploit the intrinsic feature-importance capability of KAN. Unlike post-hoc methods that explain a trained classifier ``from the outside'', KAN~2.0 introduced by~\citet{Liu2024KAN2K} includes an internal mechanism that estimates how strongly each input feature contributes to the model output based on the learned signal flow through the network. In our setting, this is particularly useful because the inputs to the SKM already correspond to semantically defined attribute posterior distributions. The resulting importance scores can therefore be interpreted directly at the attribute-value level, both globally for the attack classification task and locally for individual attack predictions.

The KAN feature-importance mechanism is based on the idea that important features induce stronger variation in the activations propagated through the network. In a KAN, each layer is computed from the previous one through learnable univariate activation functions defined on edges (Section~\ref{section:preliminary_KAN}). During training, these functions adapt to the data, and their outputs reflect how variation in the inputs is transformed and transmitted across layers. Once the model has been trained, feature-importance analysis can be performed depending on the dataset $\mathscr{D}$ fed into the model, whether the training or test set. Specifically, the importance scores are estimated over $\mathscr{D}$ by measuring the variation of node and edge activations across all samples in the dataset. Feature importance is then quantified by propagating these variation-based scores backward from the output layer to the input layer. In practice, the variation is measured by the standard deviation.

Formally, consider an $L$-layer KAN with layer widths $n^{(0)}, n^{(1)}, \ldots, n^{(L)}$, where $n^{(0)}$ is the input dimension and $n^{(L)}$ is the number of output classes. For a set of data samples $\mathscr{D}$, let $N_{l,i}(\mathscr{D})$ denote the standard deviation of the activation at node $i$ in layer $l$, and let $E_{l,i,j}(\mathscr{D})$ denote the standard deviation of the post-activation on the edge from node $i$ in layer $l$ to node $j$ in layer $l+1$. Let $S_{l,i}^{\mathbf{n}}$ and $S_{l,i,j}^{\mathbf{e}}$ denote the corresponding node and edge importance scores. Assuming equal initial importance for all output nodes, we set $S_{L,i}^{\mathbf{n}} = 1$ for $i=0,1,\ldots,n^{(L)}-1$. The scores are then propagated backward through the network according to
\begin{equation}\label{eq_edge_score}
S_{l-1,i,j}^{\mathbf{e}} = S_{l,j}^{\mathbf{n}} \cdot \frac{E_{l-1,i,j}(\mathscr{D})}{N_{l,j}(\mathscr{D})}
\end{equation}
and
\begin{equation}\label{eq_node_score}
S_{l-1,i}^{\mathbf{n}} = \sum_{j=0}^{n^{(l)}-1} S_{l-1,i,j}^{\mathbf{e}},
\end{equation}
where $l=L,L-1,\ldots,1$, $i=0,1,\ldots,n^{(l-1)}-1$, and $j=0,1,\ldots,n^{(l)}-1$. In this way, the final scores assigned to the input nodes quantify the relative contribution of the corresponding attack-attribute values to the classifier output.

\section{Experimental setup} \label{section:exp_setup}

In this section, we provide an overview of the selected dataset and describe the model configurations in detail, including the multi-task learning framework for probabilistic attribute extraction and the proposed KAN model for attack classification. To ensure reproducibility, the full implementation details and source code are publicly available\footnote{\url{https://github.com/HoangHPham/KAN-Probabilistic-Deepfake-Attribution}}.

\subsection{Data preparation} \label{section:exp_setup_dataset}

\subsubsection{ASVspoof 2019 LA dataset, closed-set protocol and metadata} \label{section:preliminary_dataset}

\begin{table*}
\centering
\caption{Indices of 50 values in each attribute of ASVspoof 2019 LA dataset \citep{WANG2020101114}. The attribute values are sequentially indexed from 1 to 50 by enumerating all rows within the column of one attribute before proceeding to the next column, starting from attribute 1 (Attr. 1) and continuing through attribute 7 (Attr. 7).}
\label{tab:asvspoof2019LA_index_attribute}
\renewcommand{\arraystretch}{1.3}
\setlength{\tabcolsep}{0.5pt}
\scriptsize
\begin{threeparttable}
\begin{tabularx}{\textwidth}{Y|Y|Y|Y|Y|Y|Y}
\hline
\makecell[Y]{\textbf{Attr. 1}\\\textbf{(inputs)}} & \makecell[Y]{\textbf{Attr. 2}\\\textbf{(processor)}} & \makecell[Y]{\textbf{Attr. 3}\\\textbf{(duration)}} & \makecell[Y]{\textbf{Attr. 4}\\\textbf{(conversion)}} & \makecell[Y]{\textbf{Attr. 5}\\\textbf{(speaker)}} & \makecell[Y]{\textbf{Attr. 6}\\\textbf{(outputs)}} & \makecell[Y]{\textbf{Attr. 7}\\\textbf{(waveform)}} \\
\hline
Text (1) & NLP (4) & HMM (10) & AR-RNN (16) & VAE (25) & MCC-F0 (30) & WaveNet (40) \\
\hline
Speech\_human (2) & WORLD (5) & FF (11) & FF (17) & One-hot (26) & MCC-F0-BAP (31) & WORLD (41) \\
\hline
Speech\_TTS (3) & LPCC/MFCC (6) & RNN (12) & CART (18) & d-vector\_RNN (27) & MFCC-F0 (32) & Concat (42) \\
\hline
{} & CNN+bi-RNN (7) & Attention (13) & VAE (19) & PLDA (28) & MCC-F0-AP (33) & SpecFiltOLA (43) \\
\hline
{} & ASR (8) & DTW (14) & GMM-UBM (20) & None (29) & LPC (34) & NeuralFilt (44) \\
\hline
{} & MFCC/i-vector (9) & None (15) & RNN (21) & {} & MCC-F0-BA (35) & Vocaine (45) \\
\hline
{} & {} & {} & AR-RNN+CNN (22) & {} & Mel-spec (36) & WaveRNN (46) \\
\hline
{} & {} & {} & Moment-match (23) & {} & F0+ling (37) & GriffinLim (47) \\
\hline
{} & {} & {} & Linear (24) & {} & MCC (38) & WaveFilt (48) \\
\hline
{} & {} & {} & {} & {} & MFCC (39) & STRAIGHT (49) \\
\hline
{} & {} & {} & {} & {} & {} & MFCCvoc (50) \\
\hline
\end{tabularx}
\end{threeparttable}
\end{table*}

The ASVspoof 2019 LA dataset \citep{WANG2020101114}, originally designed for speech deepfake detection, contains both real (bona fide) and spoofed speech data. To facilitate study of source tracing, \citet{mishra_towards} proposed two protocols: ASVspoof2019-attr-2, which follows the original ASVspoof 2019 protocol structure (but for source tracing rather than deepfake detection), and ASVspoof2019-attr-17, which focuses exclusively on spoofed speech, including 17 source generators (attacks). We adopt this larger protocol for our experiments. It partitions the data into training, development, and evaluation sets, with all 17 attacks ($A16$ and $A19$ merged into $A04$ and $A06$) present across subsets.

Despite its age, ASVspoof 2019 LA remains a valuable benchmark for source tracing research because it provides detailed metadata describing the synthesis and voice-conversion components used to generate each attack. This rich annotation enables systematic investigation of the relationships between source generators and their underlying attributes, making the dataset particularly suitable for evaluating the interpretability.

\begin{figure}
    \centering
    \includegraphics[width=1\linewidth]{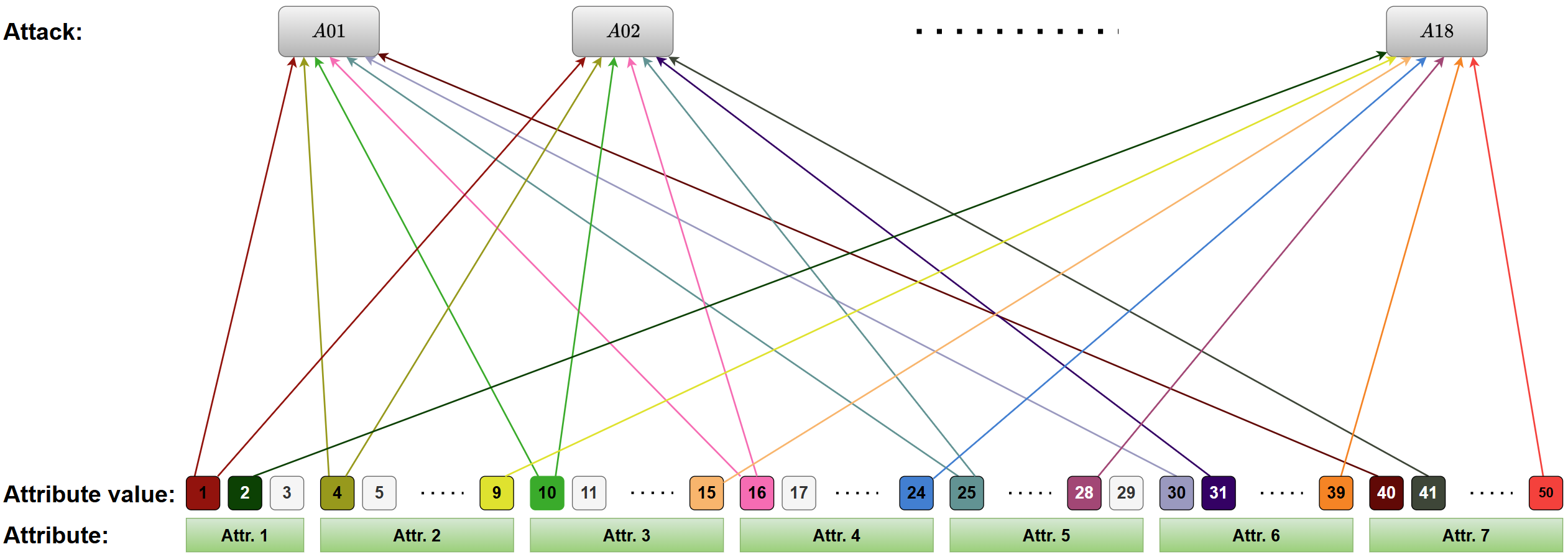}
    \caption{Structural diagram representing the relationship between the attacks and their attributes in the ASVspoof 2019 LA dataset. The dataset contains a total of 50 attribute values (\(AV\)), that we index sequentially according to their corresponding attributes. For example, \(AV01\) corresponds to ``Text (Attr. 1)'', \(AV02\) to ``Speech\_human (Attr. 1)'', \(AV03\) to ``Speech\_TTS (Attr. 1)'', \(AV04\) to ``NLP (Attr. 2)'', \(AV05\) to ``WORLD (Attr. 2)'', and so forth. For visual clarity, we only display attacks \(A01\), \(A02\), and \(A18\) as examples; \(A01\), for instance, is characterized by attribute values ``Text'', ``NLP'', ``HMM'', ``AR-RNN'', ``VAE'', ``MCC-F0'', and ``WaveNet''.}
    \label{fig:asvspoof2019LA_tree_metadata}
\end{figure}

The metadata of the ASVspoof 2019 LA dataset describes the systems used to generate each attack~\citep[Table 1]{WANG2020101114}. Each attack is produced by either a text-to-speech (TTS) or voice conversion (VC) system and is characterized by seven \emph{attributes}, corresponding to the columns of Table~\ref{tab:asvspoof2019LA_index_attribute}. The \emph{values} of these attributes, shown in the rows, specify the corresponding synthesizer subcomponents and are indexed from 1 to 50. We model the relationship between attack, or source-generator, IDs and their attribute values as a single-layer neural network, where attribute values serve as inputs and attack IDs as outputs, as illustrated in Figure~\ref{fig:asvspoof2019LA_tree_metadata}. Instead of using a conventional fully connected layer, the proposed predefined structure explicitly specifies the inter-layer connections, providing intrinsic interpretability.

\subsubsection{Preprocessing and data augmentation}

Silence removal is an essential preprocessing step, as non-speech segments carry little information and may introduce spurious patterns (\emph{shortcuts}) that degrade performance~\citep{muller21_asvspoof,Sahid2025-shortcut}. We apply \emph{silence trimming}~\citep{silence_trimming} to all utterances by removing leading and trailing silences using a hybrid approach based on \textit{short-time energy} (STE) and \textit{zero-crossing rate} (ZCR)~\citep{silence_trimming_2,silence_trimming_3}. In addition, we use \emph{RawBoost} augmentation~\citep{tak2021rawboost}, which comprises three methods: (i) linear and non-linear convolutive noise (LnL-convolutive noise), (ii) impulsive signal-dependent additive noise, and (iii) stationary signal-independent additive noise. Prior work~\citep{RawBoost_probs} has shown that combining these methods improves performance. Instead of using a fixed combination, we assign probabilities \(p_{\text{DA}}\) to all possible combinations, enabling diverse mixtures during training. Further details are provided in Appendix~\ref{app:data_prep_rawboost}.

\subsection{Model configuration} \label{section:exp_setup_model_configs}

\begin{figure}
    \centering
    \includegraphics[width=1\linewidth]{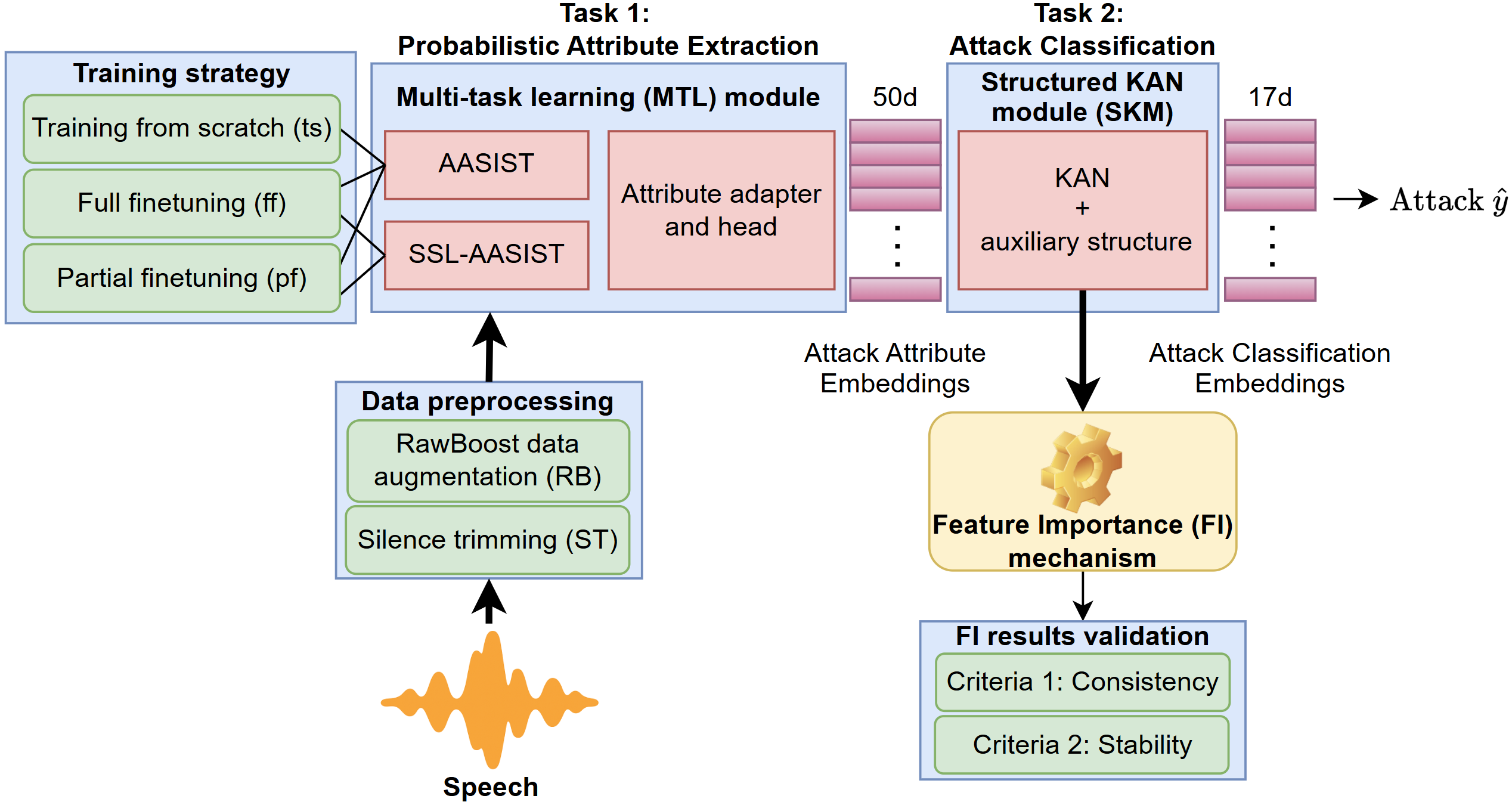}
    \caption{Overview of main components of the proposed model configurations and their roles in the experimental scenarios.}
    \label{fig:overview_experimental_scenarios}
\end{figure}

Figure~\ref{fig:overview_experimental_scenarios} summarizes the proposed model configurations and their roles in the two main experimental scenarios: (1) source tracing performance and (2) interpretability validation.

\subsubsection{Countermeasure backbones}

As described in Section~\ref{section:proposed_MTL}, our proposed source tracing model adopts deepfake detection architectures as backbones to learn shared representations for probabilistic attribute extraction. We consider three training strategies: (1) \emph{training from scratch}, (2) \emph{full fine-tuning}, and (3) \emph{partial fine-tuning}. In the first strategy, the entire model is trained from random initialization. In the latter two, the countermeasure backbone (either AASIST or SSL-AASIST) is first pretrained for the binary deepfake detection task following the original ASVspoof 2019 protocol. The pretrained weights, excluding the final classification layer, are then loaded into the shared block of the MTL module (Figure~\ref{fig:proposed_model}), after which the entire model is trained on the ASVspoof2019-attr-17 protocol. The two fine-tuning strategies differ in scope: full fine-tuning updates all parameters, whereas partial fine-tuning freezes the backbone and updates only the remaining parameters.

These strategies result in shared representations encoding different knowledge. The model trained from scratch learns only from spoofed speech in ASVspoof2019-attr-17. Full fine-tuning initializes the model with representations learned from both bona fide and spoofed speech in the original protocol, which are subsequently refined. Partial fine-tuning preserves the deepfake detection knowledge in the frozen backbone.

Table~\ref{tab:cm_backbone_performance} reports the deepfake detection performance of the pretrained AASIST and SSL-AASIST models on the ASVspoof2019-attr-2 protocol, evaluated with the \emph{equal error rate} (EER)~\citep{EER}. We adopt the architectures and configurations from the original papers~\citep{Jung2021AASIST, tak2022automatic}. For SSL-AASIST, the pretrained wav2vec 2.0 XLS-R (0.3B) model~\citep{babu2022xlsr} is fine-tuned on ASVspoof2019-attr-2; training from scratch is omitted because it would require data and computational resources unavailable in this study.

\begin{table}[htbp]
\centering
\caption{Performance of the pretrained countermeasure (CM) models (AASIST and SSL-AASIST) on the deepfake detection task using the ASVspoof2019-attr-2 protocol. Pretrained weights are loaded into the shared module of the proposed model.}
\label{tab:cm_backbone_performance}
\renewcommand{\arraystretch}{1.5}
\begin{tabular}{l|c}
\hline
\multicolumn{1}{c|}{\textbf{CM model}} & 
\multicolumn{1}{c}{\textbf{EER (\%)}} \\
\hline
AASIST & 2.82 \\
\hline
SSL-AASIST & 0.29 \\
\hline
\end{tabular}
\end{table}

\subsubsection{Probabilistic attribute extractors trained using multi-task learning}

The multi-task learning (MTL) module is used to train the probabilistic attribute extractors (Section~\ref{section:proposed_MTL}). It consists of shared layers and attribute-specific layers. The shared layers correspond to one of the backbones presented above, while each attribute-specific block comprises an attribute adapter and an attribute head. The attribute adapter follows the Houlsby adapter design~\citep{houlsby2019parameter}, and the attribute head is a simple linear layer. All seven attribute-specific extractors share the same adapter architecture: layer normalization is first applied to stabilize the shared representations, followed by a bottleneck structure with down-projection and up-projection layers, GELU activation~\citep{Gelu} for nonlinearity, and dropout for regularization. A residual connection preserves shared information and ensures stable gradient flow, and an output projection layer maps the adapted representations to the attribute-specific feature space. Each attribute head contains a number of output neurons equal to the number of values in its corresponding attribute; specifically, attributes 1 through 7 have 3, 6, 6, 9, 5, 10, and 11 values, respectively, as indicated in  Table~\ref{tab:asvspoof2019LA_index_attribute}.

\subsubsection{Structured KAN module (SKM)}

The objective of the SKM is to classify 17 attacks using the 50-dimensional logit vector produced by the bank of probabilistic attribute extractors. It is a KAN (Section~\ref{section:preliminary_KAN}) with a single hidden layer consisting of $50 \times 17$ learnable univariate functions. Each function uses 5 grid intervals ($G=5$) and a piecewise polynomial of order 3 (degree-2 B-spline). To embed the auxiliary structure derived from the metadata of the ASVspoof 2019 LA dataset into the KAN, we manually configure its edges---namely, only the edges corresponding to valid attribute values for each attack are retained, following the hierarchy in Figure~\ref{fig:asvspoof2019LA_tree_metadata}. Further details are provided in Appendix~\ref{app:exp_setup_KAN_aux}.

\subsection{Performance metrics}
\label{section:exp_setup_metrics}

We evaluate the performance of both probabilistic attribute extraction and attack classification using \emph{balanced accuracy} (BAcc)\footnote{\url{https://scikit-learn.org/dev/modules/model_evaluation.html\#balanced-accuracy-score}} and \emph{equal error rate} (EER)~\citep{EER}. Balanced accuracy is preferred over standard accuracy because the evaluation set exhibits an uneven distribution of samples across attribute values and attack classes. It is defined as the average recall over all classes:
\begin{equation}
\mathrm{BAcc}
=
\frac{1}{C}
\sum_{c=1}^{C}
\frac{\text{TP}(c)}
     {\text{TP}(c)+\text{FN}(c)}
\end{equation}
where $C$ denotes the number of classes, and $\text{TP}(c)$ and $\text{FN}(c)$ denote the numbers of true positives and false negatives for class $c$, respectively. A higher BAcc indicates better classification performance.

In addition to BAcc, we report EER, which is widely used in speech processing and biometric recognition tasks. Since both probabilistic attribute extraction and attack classification are formulated as multi-class classification problems, EER is computed using a one-versus-rest strategy following~\citep{mishra_towards}. For probabilistic attribute extraction, the ground-truth value of an attribute is treated as the target class (positive class), while all remaining values of the same attribute are treated as non-target classes (negative class). The distributions of target and non-target scores are then used to compute the EER. Similarly, for attack classification, the ground-truth attack is considered the target class and the remaining attacks are considered non-target classes. Lower EER indicates better discrimination between target and non-target classes.

All reported results are obtained on the evaluation set of the ASVspoof2019-attr-17 protocol. For attack classification, the attack label predicted by the model is directly used to compute the evaluation metrics. For probabilistic attribute extraction, BAcc and EER are computed independently for each attribute.

\section{Experimental results}

\subsection{Experiment 1: Source tracing performance}

\subsubsection{Baseline model}

We first reproduce the baseline results reported in~\citep{manasi_explainable, mishra_towards} using the two-stage training framework to establish the reference performance for comparison. Different from these two prior studies, our reproduced setup additionally incorporates the silence trimming and RawBoost augmentations described above. Among all reproduced configurations, the best performance is achieved using probabilistic attribute extractors trained on latent features extracted from the AASIST model. For attack classification, the extracted probabilistic embeddings are used as inputs to a logistic regression (LR) back-end.

For probabilistic attribute extraction, the best performance is achieved for Attribute~1, with a balanced accuracy of 91.59\% and an EER of 3.16\%. The remaining attributes also show stable performance, with balanced accuracies ranging from 83.82\% to 89.51\% and EERs between 3.22\% and 5.66\%. Using concatenated probabilistic attribute embeddings, the LR classifier achieves a balanced accuracy of 84.37\% and an EER of 3.35\%, yielding the best baseline performance for attack classification. A detailed comparison of the reproduced baseline configurations and their corresponding performance is provided in Appendix~\ref{app:baseline_results}.

\subsubsection{Proposed model}

\begin{table}[htbp]
\centering
\caption{Performance of our proposed source tracing model (end-to-end workflow of countermeasure based multi-task learning and Kolmogorov-Arnold Network) on ASVspoof2019-attr-17 protocol. The results on evaluation set are reported as \(<\mathrm{balanced~accuracy}>\) (\%) / \(<\mathrm{equal~error~rate}>\) (\%). The abbreviations in the table represent the following: BAcc (balanced accuracy), ST (silence trimming), RB (RawBoost), ts (training from scratch), ff (full finetuning), pf (partial finetuning), MTL (multi-task learning), SKM (structured KAN module).}
\label{tab:exp_proposed_sourcetracing}
\renewcommand{\arraystretch}{2.2}
\scriptsize
\begin{threeparttable}
\begin{tabular}{c|c|c|c|c|c|c|c|c}
\Xhline{1.5pt}
\multicolumn{8}{c|}{MTL} & SKM \\
\hline
\makecell[c]{Countermeasure \\ backbone} & 
\multicolumn{7}{c|}{\textbf{Probabilistic attribute extraction}} & \textbf{Attack classification} \\
\hline
& \multicolumn{7}{c|}{BAcc (\%) / EER (\%)} & \multirow{2}{*}{BAcc (\%) / EER (\%)} \\
\cline{2-8}
& Attr. 1 & Attr. 2 & Attr. 3 & Attr. 4 & Attr. 5 & Attr. 6 & Attr. 7 &  \\
\Xhline{1.5pt}
\texttt{ST\_RB\_AASIST\_ts}
    & \makecell[l]{99.98~/\\0.02} 
    & \makecell[l]{99.72~/\\0.17}
    & \makecell[l]{99.54~/\\0.17}
    & \makecell[l]{99.68~/\\0.13}
    & \makecell[l]{99.61~/\\0.21}
    & \makecell[l]{99.37~/\\0.15}
    & \makecell[l]{99.60~/\\0.14}
    & 99.53~/~0.12 \\
    \cline{1-9}
\texttt{ST\_RB\_AASIST\_ff}
    & \makecell[l]{99.99~/\\0.01}
    & \makecell[l]{99.80~/\\0.13}
    & \makecell[l]{99.60~/\\0.19}
    & \makecell[l]{99.68~/\\0.12}
    & \makecell[l]{99.66~/\\0.20}
    & \makecell[l]{99.48~/\\0.15}
    & \makecell[l]{99.61~/\\0.13}
    & 99.61~/~0.11 \\
    \cline{1-9}
\texttt{ST\_RB\_AASIST\_pf}
    & \makecell[l]{91.57~/\\3.69}
    & \makecell[l]{80.79~/\\4.87}
    & \makecell[l]{83.27~/\\5.14}
    & \makecell[l]{85.64~/\\3.91}
    & \makecell[l]{79.81~/\\6.98}
    & \makecell[l]{82.20~/\\4.43}
    & \makecell[l]{82.58~/\\4.50}
    & 81.96~/~3.64 \\
    \cline{1-9}
\Xcline{1-9}{1.5pt}
\texttt{ST\_RB\_SSL-AASIST\_ff}
    & \makecell[l]{99.92~/\\0.07}
    & \makecell[l]{99.76~/\\0.12}
    & \makecell[l]{99.68~/\\0.14}
    & \makecell[l]{99.74~/\\0.12}
    & \makecell[l]{99.68~/\\0.16}
    & \makecell[l]{99.59~/\\0.15}
    & \makecell[l]{99.63~/\\0.14}
    & 99.64~/~0.11 \\
    \cline{1-9}
\texttt{ST\_RB\_SSL-AASIST\_pf}
    & \makecell[l]{69.66~/\\15.07}
    & \makecell[l]{50.38~/\\13.32}
    & \makecell[l]{54.30~/\\15.31}
    & \makecell[l]{62.41~/\\11.82}
    & \makecell[l]{59.75~/\\15.70}
    & \makecell[l]{57.54~/\\12.12}
    & \makecell[l]{61.31~/\\12.84}
    & 61.71~/~9.86 \\
    \cline{1-9}
\Xhline{1.5pt}
\end{tabular}
\end{threeparttable}
\end{table}

Building on the extended reproduced baseline, which incorporates both silence trimming and RawBoost, we next present the results of the proposed system, which further integrates multitask learning for probabilistic attribute extraction and KAN-based attack classification. The results in Table~\ref{tab:exp_proposed_sourcetracing} indicate that the proposed end-to-end model consistently outperforms the baselines on both tasks.

When using AASIST as the backbone, both the training-from-scratch and full-finetuning configurations (i.e., \texttt{ST\_RB\_AASIST\_ts} and \texttt{ST\_RB\_AASIST\_ff}) achieve balanced accuracies above 99\% across all attribute extractors, despite differences in the number of predicted attribute values. This indicates that the multi-task learning (MTL) architecture effectively captures shared representations while maintaining strong attribute-specific performance. In the attack classification task, the model reaches 99.53\% and 99.61\% balanced accuracy and 0.12\% and 0.11\% EER for \texttt{AASIST\_ts} and \texttt{AASIST\_ff}, respectively. This suggests that the model trained from scratch achieves performance comparable to that of the fully fine-tuned model, despite the latter being trained for more epochs and using both bona fide and spoofed speech. In contrast, partial finetuning significantly degrades performance, yielding balanced accuracies between 79.81\% and 91.57\% (6.98\%--3.69\% EER) for attribute extraction and 81.96\% (EER: 3.64\%) for attack classification.

We observe similar trends when using SSL-AASIST as the backbone. With full fine-tuning (\texttt{SSL-AASIST\_ff}), the model achieves highly stable performance across the attribute extraction tasks, with balanced accuracies ranging from 99.59\% to 99.92\% and EERs from 0.15\% to 0.07\%. For attack classification, this model achieves 99.64\% balanced accuracy and 0.11\% EER. In contrast, partial fine-tuning of SSL-AASIST yields the lowest performance among all configurations, with attribute extraction accuracies ranging from 50.38\% to 69.66\% and EERs from 15.70\% to 11.82\%. The corresponding attack classification performance is 61.71\% balanced accuracy and 9.86\% EER. These results further indicate that training from scratch or full fine-tuning is more suitable than partial fine-tuning when using countermeasure models as backbones. For an easier comparison, we compare the strongest baseline, based on AASIST features with an LR classifier, with the best proposed configuration, \texttt{ST\_RB\_SSL-AASIST\_ff}. As shown in Table~\ref{tab:comparison_baseline_ppm}, the proposed model substantially outperforms the baseline in both attribute extraction and attack classification.

\begin{table}[htbp]
\centering

\caption{Comparison between the best proposed model, \texttt{ST\_RB\_SSL-AASIST\_ff} (multi-task learning module using full-finetuned SSL-AASIST and structured KAN module) and baselines (AASIST-based feature extractor combined with logistic regression) for probabilistic attribute extraction and attack classification on the ASVspoof2019-attr-17 protocol. The reported baseline results include those originally reported by \citet{mishra_towards} and the corresponding baseline reproduced in this work using the same preprocessing pipeline as the proposed system (silence trimming and Rawboost augmentation). Results are presented as \(<\mathrm{balanced~accuracy}>\) (BAcc, \%) / \(<\mathrm{equal~error~rate}>\) (EER, \%). A dash ("--") indicates that the corresponding metric was not reported in the original study.}
\label{tab:comparison_baseline_ppm}
\renewcommand{\arraystretch}{2.2}
\scriptsize
\begin{threeparttable}
\begin{tabular}{c|c|c|c|c|c|c|c|c}
\Xhline{1.5pt}
\makecell[c]{Model} & 
\multicolumn{7}{c|}{\textbf{Probabilistic attribute extraction}} & \textbf{Attack classification} \\
\hline
& \multicolumn{7}{c|}{BAcc (\%) / EER (\%)} & \multirow{2}{*}{BAcc (\%) / EER (\%)} \\
\cline{2-8}
& Attr. 1 & Attr. 2 & Attr. 3 & Attr. 4 & Attr. 5 & Attr. 6 & Attr. 7 &  \\
\Xhline{1.5pt}
baseline model (reported)
    & \makecell[l]{--~/\\2.0}
    & \makecell[l]{--~/\\3.2}
    & \makecell[l]{--~/\\3.3}
    & \makecell[l]{--~/\\2.3}
    & \makecell[l]{--~/\\4.2}
    & \makecell[l]{--~/\\3.0}
    & \makecell[l]{--~/\\2.9}
    & 90.23~/~2.07 \\
    \cline{1-9}
baseline model (reproduced)
    & \makecell[l]{91.59~/\\3.16}
    & \makecell[l]{87.16~/\\3.92}
    & \makecell[l]{85.26~/\\4.58}
    & \makecell[l]{89.51~/\\3.22}
    & \makecell[l]{86.25~/\\5.66}
    & \makecell[l]{83.82~/\\4.00}
    & \makecell[l]{86.06~/\\3.96}
    & 84.37~/~3.35 \\
    \cline{1-9}
proposed model
    & \makecell[l]{\textbf{99.92~/}\\ \textbf{0.07}}
    & \makecell[l]{\textbf{99.76~/}\\ \textbf{0.12}}
    & \makecell[l]{\textbf{99.68~/}\\ \textbf{0.14}}
    & \makecell[l]{\textbf{99.74~/}\\ \textbf{0.12}}
    & \makecell[l]{\textbf{99.68~/}\\ \textbf{0.16}}
    & \makecell[l]{\textbf{99.59~/}\\ \textbf{0.15}}
    & \makecell[l]{\textbf{99.63~/}\\ \textbf{0.14}}
    & \textbf{99.64~/~0.11} \\
    \cline{1-9}
\Xhline{1.5pt}
\end{tabular}
\end{threeparttable}
\end{table}

\subsubsection{Ablation study: KAN without auxiliary model structure}

With the main classification experiments completed, we now assess the contribution of the metadata-derived auxiliary structure in the proposed KAN module. We conduct an ablation study in which this auxiliary structure, derived from the ASVspoof 2019 LA metadata, is removed from the KAN module. Specifically, instead of manually defining the connections between input features and activation functions, we adopt a fully connected KAN architecture in which each of the 50 input nodes is connected to 17 hidden activation functions corresponding to the output units (one per attack). Table~\ref{tab:exp_ablation_KAN_without_MS} summarizes the results obtained using AASIST as the backbone under the three training strategies described above.

The results indicate that the ablated model and the proposed model achieve highly similar performance in both attribute extraction and attack classification. For attack classification, the fully connected configuration with \texttt{AASIST\_ff} achieves a balanced accuracy of 99.65\%, compared to 99.61\% obtained by the proposed model using SKM. The corresponding results for \texttt{AASIST\_ts} and \texttt{AASIST\_pf} are 99.58\% and 82.41\%, respectively, which are also very close to those obtained using SKM.

In general, the findings suggest that incorporating the metadata-derived auxiliary structure does not substantially affect classification performance. The fully connected KAN and the proposed structured KAN provide comparable results across all training strategies. Consequently, the choice between the two architectures may be primarily guided by factors other than predictive performance. In the present work, the auxiliary structure is motivated by its ability to explicitly encode known relationships between attributes and attacks, thereby providing an interpretable model structure. At the same time, the results indicate that a conventional fully connected KAN remains a viable alternative when interpretability is not a primary consideration.

\begin{table}[htbp]
\centering
\caption{Experimental results of the ablation study - proposed model without auxiliary model structure in KAN module. The results on evaluation set of ASVspoof2019-attr-17 protocol are shown in the form of \(<\mathrm{balanced~accuracy}>\) (\%). The abbreviations in the table represent the following: BAcc (balanced accuracy), ST (silence trimming), RB (RawBoost), ts (training from scratch), ff (full finetuning), pf (partial finetuning), MTL (multi-task learning), KAN w/o MS (KAN without auxiliary model structure).}
\label{tab:exp_ablation_KAN_without_MS}
\renewcommand{\arraystretch}{2.2}
\scriptsize
\begin{threeparttable}
\begin{tabular}{c|c|c|c|c|c|c|c|c}
\Xhline{1.5pt}
\multicolumn{8}{c|}{MTL} & KAN \textbf{w/o} aux MS \\
\hline
\makecell[c]{Countermeasure \\ backbone} & 
\multicolumn{7}{c|}{\textbf{Probabilistic attribute extraction}} & \textbf{Attack classification} \\
\hline
& \multicolumn{7}{c|}{BAcc (\%)} & \multirow{2}{*}{BAcc (\%)} \\
\cline{2-8}
& Attr. 1 & Attr. 2 & Attr. 3 & Attr. 4 & Attr. 5 & Attr. 6 & Attr. 7 &  \\
\Xhline{1.5pt}
\texttt{ST\_RB\_AASIST\_ts}
    & 99.99
    & 99.72
    & 99.55
    & 99.66
    & 99.59
    & 99.50
    & 99.57
    & 99.58 \\
    \cline{1-9}
\texttt{ST\_RB\_AASIST\_ff}
    & 99.99
    & 99.78
    & 99.63
    & 99.69
    & 99.66
    & 99.53
    & 99.64
    & 99.65 \\
    \cline{1-9}
\texttt{ST\_RB\_AASIST\_pf}
    & 91.20
    & 81.05
    & 83.21
    & 85.60
    & 80.08
    & 81.75
    & 82.57
    & 82.41 \\
    \cline{1-9}
\Xhline{1.5pt}
\end{tabular}
\end{threeparttable}
\end{table}

\subsection{Experiment 2: Feature importance analysis and validation}

\subsubsection{Global and local feature importance at model level and per-class level}\label{section:exp_res_global_local_fi}

We now examine feature importance (FI) within the structured KAN module, where the inputs correspond to the 50 attribute values and the outputs correspond to the 17 attacks. We aim to quantify the contribution of attack attributes to the predicted attacks. The FI results are categorized along two criteria:

\begin{itemize}
    \item \textbf{Amount of data observed by the model:}
    \begin{itemize}
        \item \emph{\textbf{Global FI}} reflects feature importance over the entire dataset.
        \item \emph{\textbf{Local FI}} captures feature importance within a specific batch of samples.
    \end{itemize}

    \item \textbf{Target of interpretation:}
    \begin{itemize}
        \item \emph{\textbf{Model-level FI}} measures the contribution of each feature to the overall system performance.
        \item \emph{\textbf{Per-class FI}} highlights the importance of features in predicting a specific attack.
    \end{itemize}
\end{itemize}

In other words, model-level FI focuses on the importance scores of all 50 input attack attribute values, whereas per-class FI focuses on the seven attributes associated with each attack. For consistency, we assess feature importance produced by the SKM using the \texttt{ST\_RB\_AASIST\_MTL\_SKM\_ts} configuration, which uses AASIST as the backbone and is trained from scratch. The evaluation set of the ASVspoof2019-attr-17 protocol is used for all interpretability experiments.

\begin{figure}
    \centering
    \includegraphics[width=1\linewidth]{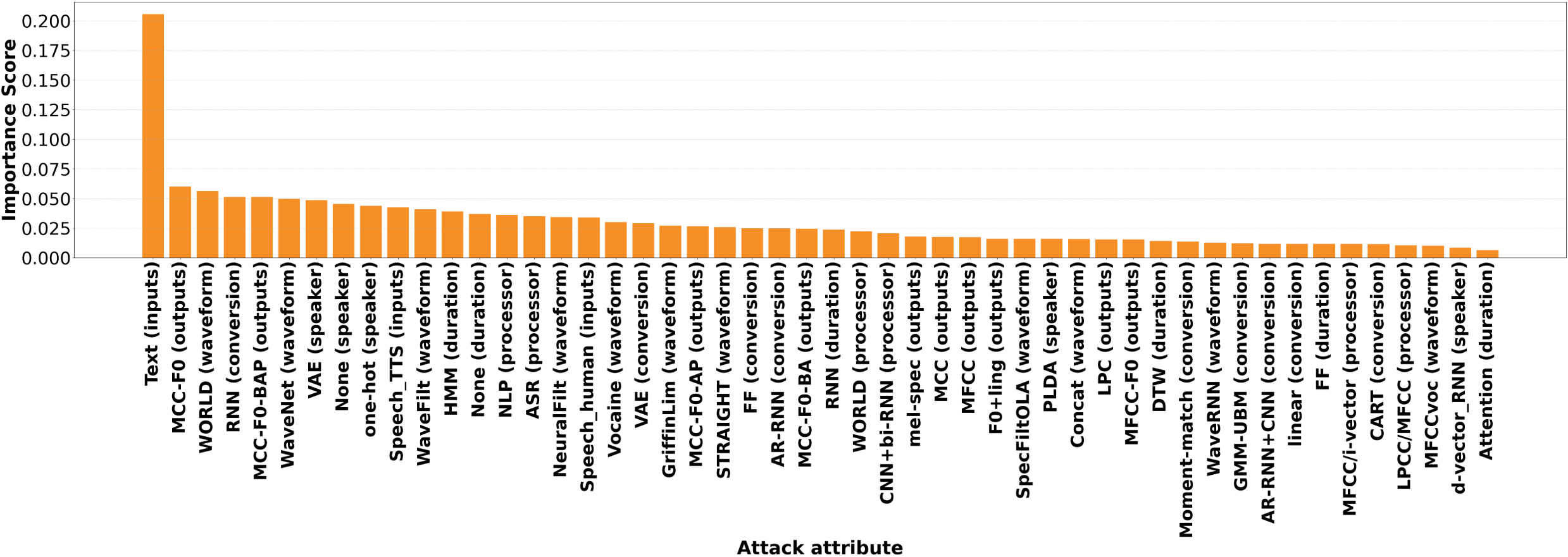}\
    \caption{Global importance scores of the 50 attack attribute values at the model-level of the attack classification task. Attribute values are ranked in descending order of importance.}
    \label{fig:global_FI_ranks_model_level}
\end{figure}

\begingroup
\setlength{\tabcolsep}{0pt} 
\renewcommand{\arraystretch}{0.1} 
\newcolumntype{P}{>{\centering\arraybackslash}p{2.2cm}}
\begin{table}[t]
\centering
\caption{Global importance ranking of each attribute value at the per-class level of the attack classification task. The attribute values are ranked in descending order of importance, with each row corresponding to a specific attack class. Attacks A04 and A16 share identical attributes and are reported jointly; the same applies to A06 and A19.}
\label{tab:global_FI_ranks_per_class_level}
\renewcommand{\arraystretch}{2}
\tiny
\begin{tabular}{c|*{6}{P|}P}
\hline
\multirow{2}{*}{\textbf{Attack}} & \multicolumn{7}{c}{\textbf{Rank}} \\
\cline{2-8}
& \textbf{1} & \textbf{2} & \textbf{3} & \textbf{4} & \textbf{5} & \textbf{6} & \textbf{7} \\
\hline
\textbf{A01} & VAE (speaker) & WaveNet (waveform) & MCC-F0 (outputs) & HMM (duration) & AR-RNN (conversion) & Text (inputs) & NLP (processor) \\
\hline
\textbf{A02} & Text (inputs) & VAE (speaker) & HMM (duration) & MCC-F0-BAP (outputs) & WORLD (waveform) & AR-RNN (conversion) & NLP (processor) \\
\hline
\textbf{A03} & FF (conversion) & MCC-F0-BAP (outputs) & Text (inputs) & FF (duration) & NLP (processor) & WORLD (waveform) & one-hot (speaker) \\
\hline
\textbf{A04(16)} & Text (inputs) & Concat (waveform) & MFCC-F0 (outputs) & CART (conversion) & None (speaker) & None (duration) & NLP (processor) \\
\hline
\textbf{A05} & MCC-F0-AP (outputs) & WORLD (waveform) & VAE (conversion) & WORLD (processor) & Speech\_human (inputs) & None (duration) & one-hot (speaker) \\
\hline
\textbf{A06(19)} & SpecFiltOLA (waveform) & LPC (outputs) & GMM-UBM (conversion) & LPCC/MFCC (processor) & None (speaker) & Speech\_human (inputs) & None (duration) \\
\hline
\textbf{A07} & WORLD (waveform) & MCC-F0-BA (outputs) & Text (inputs) & RNN (conversion) & RNN (duration) & NLP (processor) & one-hot (speaker) \\
\hline
\textbf{A08} & NeuralFilt (waveform) & HMM (duration) & AR-RNN (conversion) & NLP (processor) & MCC-F0 (outputs) & Text (inputs) & one-hot (speaker) \\
\hline
\textbf{A09} & Vocain (waveform) & Text (inputs) & MCC-F0 (outputs) & RNN (conversion) & one-hot (speaker) & RNN (duration) & NLP (processor) \\
\hline
\textbf{A10} & Text (inputs) & WaveRNN (waveform) & mel-spec (outputs) & CNN+bi-RNN (processor) & AR-RNN+CNN (conversion) & d-vector\_RNN (speaker) & Attention (duration) \\
\hline
\textbf{A11} & GriffinLim (waveform) & Text (inputs) & CNN+bi-RNN (processor) & mel-spec (outputs) & d-vector\_RNN (speaker) & AR-RNN+CNN (conversion) & Attention (duration) \\
\hline
\textbf{A12} & Text (inputs) & WaveNet (waveform) & F0+ling (outputs) & one-hot (speaker) & RNN (conversion) & NLP (processor) & RNN (duration) \\
\hline
\textbf{A13} & MCC (outputs) & DTW (duration) & Moment-match (conversion) & None (speaker) & Speech\_TTS (inputs) & WaveFilt (waveform) & WORLD (processor) \\
\hline
\textbf{A14} & STRAIGHT (waveform) & MCC-F0-BAP (outputs) & ASR (processor) & Speech\_TTS (inputs) & RNN (conversion) & None (speaker) & None (duration) \\
\hline
\textbf{A15} & ASR (processor) & Speech\_TTS (inputs) & WaveNet (waveform) & MCC-F0 (outputs) & None (speaker) & RNN (conversion) & None (duration) \\
\hline
\textbf{A17} & WaveFilt (waveform) & VAE (conversion) & Speech\_human (inputs) & MCC-F0 (outputs) & one-hot (speaker) & None (duration) & WORLD (processor) \\
\hline
\textbf{A18} & MFCC (outputs) & PLDA (speaker) & linear (conversion) & MFCC/i-vector (processor) & MFCCvoc (waveform) & None (duration) & Speech\_human (inputs) \\
\hline
\end{tabular}
\end{table}
\endgroup

We first present the rankings of global feature importance at the model level and per-class level in Figure~\ref{fig:global_FI_ranks_model_level} and Table~\ref{tab:global_FI_ranks_per_class_level}, respectively. The feature importance scores are computed using the KAN-based mechanism described in Section~\ref{section:proposed_fi_mechanism}, which propagates activation variability through the network to quantify the contribution of features.

At the model level, the results show that the value \emph{text} of the \emph{inputs} attribute dominates the importance scores, reflecting its strong influence on the decision-making process, whereas \emph{attention (duration)} contributes the least. Attribute values from the \emph{conversion} and \emph{processor} attributes generally exhibit relatively low importance. At the per-class level, in contrast, values from the \emph{inputs}, \emph{waveform}, and \emph{outputs} attributes frequently rank among the most important features for predicting specific attacks, while values from the \emph{duration} and \emph{processor} attributes typically appear near the bottom of the rankings.

These findings are consistent with those reported by \citet{mishra_towards}, who investigated attack-attribute relevance using a model-agnostic SHAP-based attribution framework~\citep{Shapley}. Despite the different methodology, both approaches consistently identify system input type (e.g., speech in VC systems or text in TTS systems) and waveform generation as key discriminative factors for attack classification.

These patterns suggest that the classifier relies primarily on high-level properties of the spoofing pipeline rather than on secondary implementation details. In particular, the prominence of the \emph{inputs} attribute indicates that the generation paradigm itself provides a strong cue for distinguishing attacks. The per-class rankings further indicate that specific attacks are characterized by distinct combinations of input type, waveform generation, and output representation. Overall, these results support the interpretability of the structured KAN module by showing that its decisions are driven by technologically meaningful attributes of the spoofing systems.

\subsubsection{Criteria 1: Consistency (KAN's built-in FI vs. SHAP values)}

\begin{figure}
    \centering
    \includegraphics[width=0.5\linewidth]{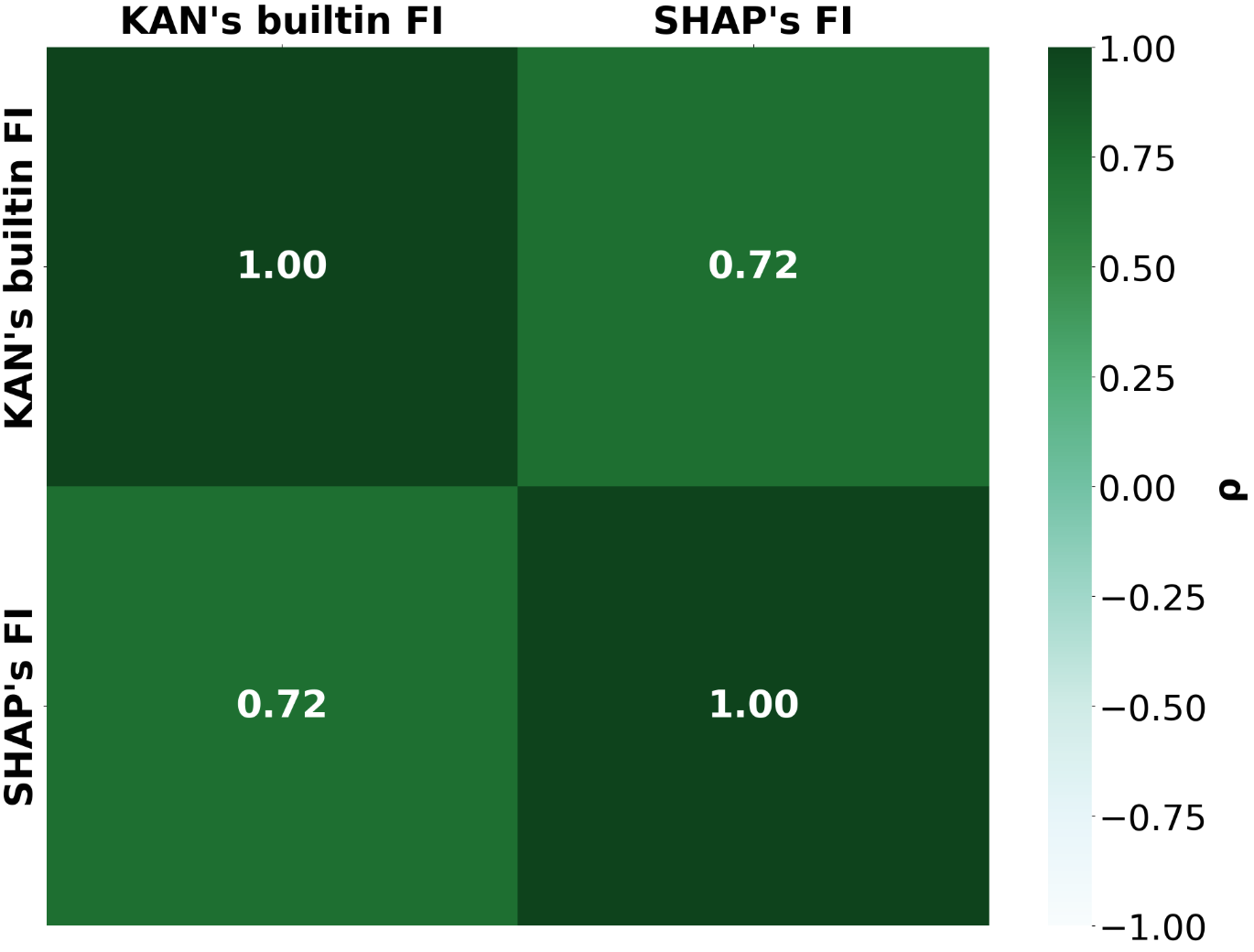}
    \caption{Spearman rank correlation of global feature importance at model-level using KAN's built-in FI mechanism and SHAP method.}
    \label{fig:exp_res_FIeval_consistency_diffFImethods_modelLevel}
\end{figure}

\begin{figure}
    \centering
    \includegraphics[width=1\linewidth]{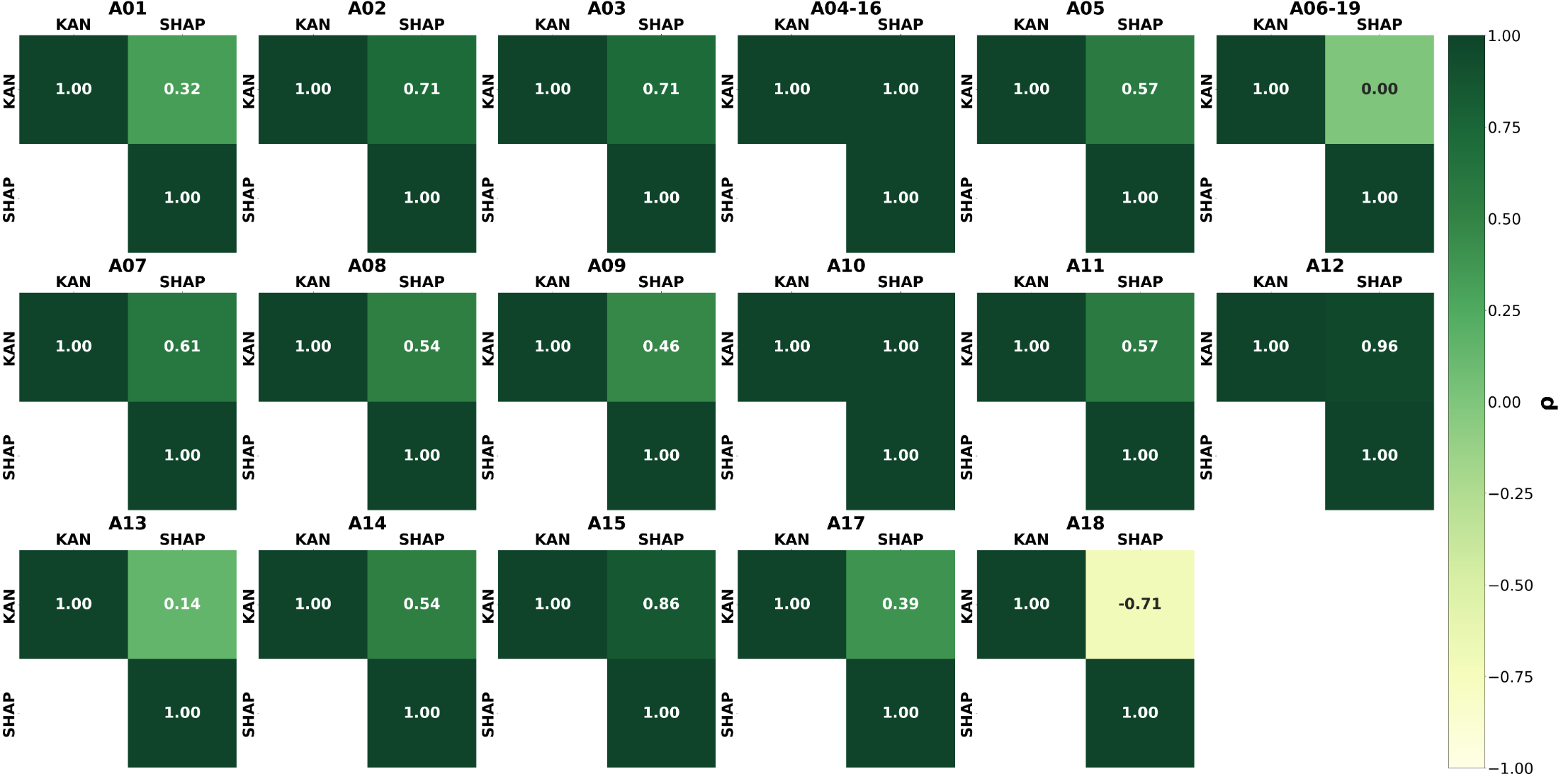}
    \caption{Spearman rank correlation of global feature importance at per-class level using KAN's built-in FI mechanism and SHAP method.}
    \label{fig:exp_res_FIeval_consistency_diffFImethods_perClass}
\end{figure}

To evaluate \emph{consistency}, we examine whether the FI results are significantly different between different interpretability methods. Specifically, we compare the FI rankings obtained from KAN's built-in feature importance mechanism with those derived from a widely used post-hoc method, namely \emph{SHapley Additive exPlanations} (SHAP)~\citep{Shapley}. We perform the analysis for global feature importance at both the model and per-class levels. To quantify agreement between KAN- and SHAP-based rankings, we use the Spearman rank correlation coefficient~\citep{Spearman_1,Spearman_2,Spearman_3}. For global FI, both methods produce one importance value per feature over the entire evaluation set, yielding two 50-dimensional vectors corresponding to the 50 attribute values. The Spearman correlation is computed between these vectors. For per-class analysis, the same procedure is applied separately for each attack class using 7-dimensional vectors corresponding to the 7 attributes. Figures~\ref{fig:exp_res_FIeval_consistency_diffFImethods_modelLevel} and~\ref{fig:exp_res_FIeval_consistency_diffFImethods_perClass} present the results for the model-level and per-class analyses, respectively.

The results reveal that the feature importance (FI) rankings produced by KAN's built-in mechanism largely agree with those obtained from SHAP. At the model level, the two methods exhibit a strong correlation, with a Spearman coefficient of 0.72. We observe a similar trend at the per-class level, where most attacks show positive rank correlations. Lower correlations (below 0.5) occur for \(A01\), \(A06(19)\), \(A09\), \(A13\), and \(A17\), while \(A18\) is the only case with a negative coefficient (\(-0.71\)). Overall, these results indicate consistency between the two methods and suggest that KAN's built-in FI mechanism can reliably capture feature importance in a manner comparable to the widely used SHAP approach.

\subsubsection{Criteria 2: Stability (with respect to input data size)} \label{section:exp_results_FI_eval_stability}

In the context of model explainability, \emph{stability} requires that explanation results remain reasonably consistent when the same model processes different input data. In our experiment, in particular, we assess the stability of the FI mechanism across different batch sizes by comparing importance scores computed from mini-batches and then aggregated (local FI) with those obtained from the full dataset (global FI). We compute FI using batch sizes of 1, 8, 32, 64, 256, and full batch. In this context, a batch denotes a set of evaluation samples processed jointly to compute feature importance, as defined in Section~\ref{section:proposed_fi_mechanism}, and is unrelated to model training mini-batches. Smaller batch sizes are expected to introduce higher variance in the estimated FI scores and may therefore lead to less stable rankings. We perform both global and local FI analyses at the model and per-class levels. To quantify stability, we compare the resulting rankings using the Spearman rank correlation coefficient.

\begin{figure}
    \centering
    \includegraphics[width=0.5\linewidth]{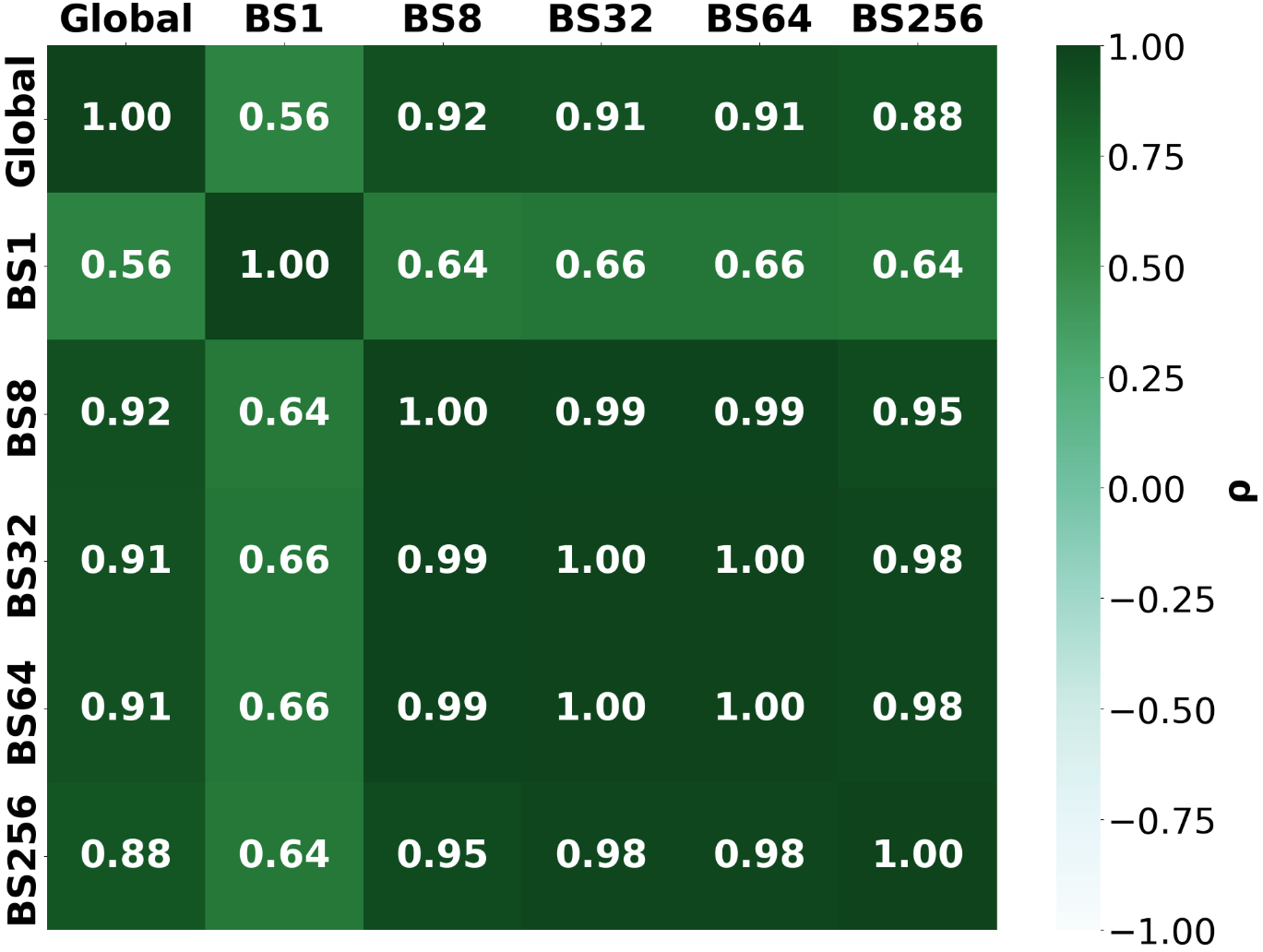}
    \caption{Spearman rank correlation of global and local feature importance at model-level.}
    \label{fig:FI_stability_different_BS_whole_system}
\end{figure}

\begin{figure}
    \centering
    \includegraphics[width=1\linewidth]{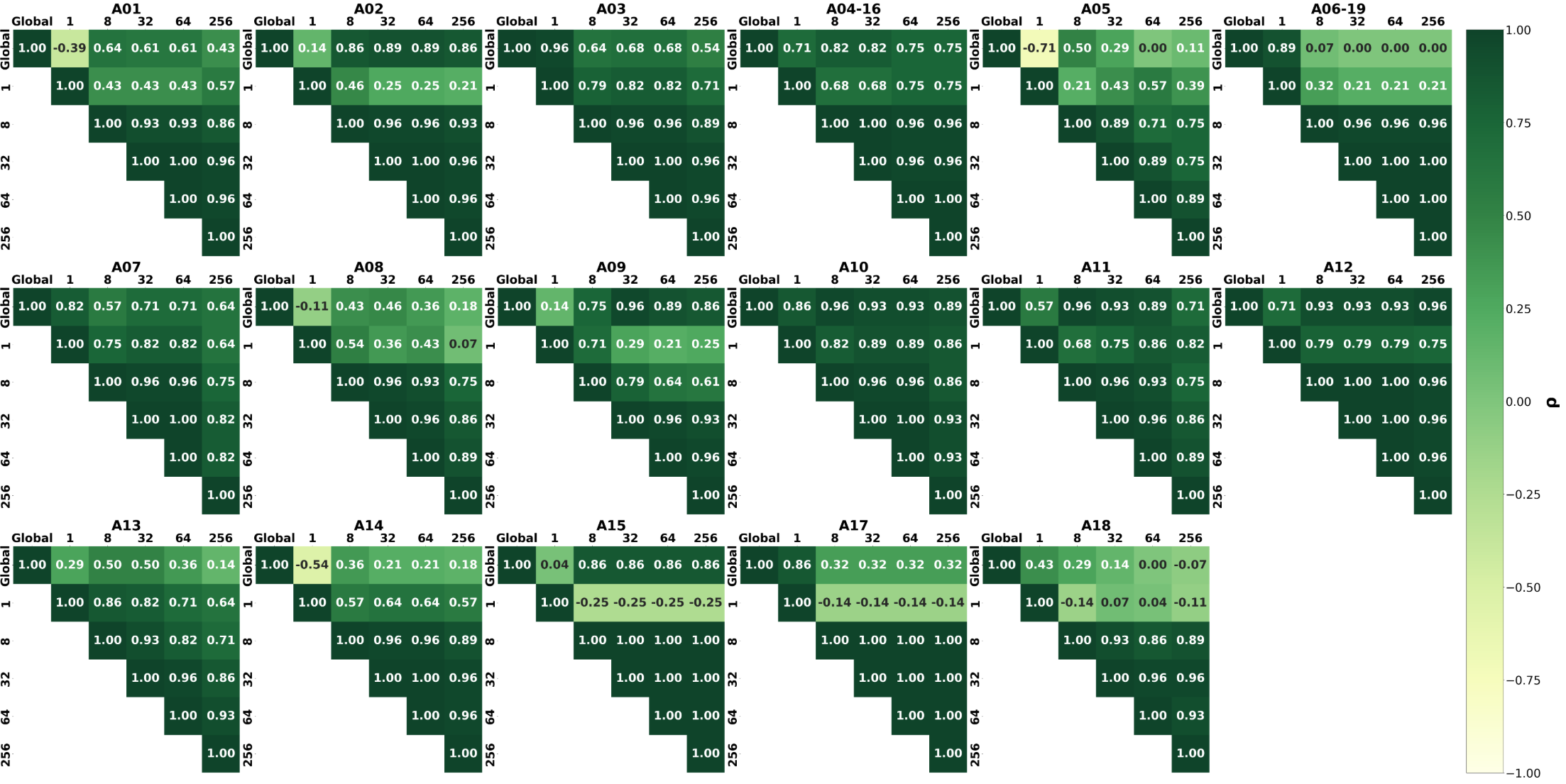}
    \caption{Spearman rank correlation of global and local feature importance at per-class level.}
    \label{fig:FI_stability_different_BS_per_attack}
\end{figure}

Figure~\ref{fig:FI_stability_different_BS_whole_system} shows that the Spearman rank correlation between global and local feature importance (FI) at the model level remains high for moderate to large batch sizes (8, 32, 64, and 256), ranging from 0.88 to 0.92. Correlations among the corresponding local FI estimates are similarly stable, ranging from 0.95 (between batch sizes 8 and 256) to 1.00, indicating strong consistency across these settings. In contrast, batch size 1 yields substantially lower correlations, both with global FI (0.56) and with the other local FI estimates (0.64--0.66), as one might expect. Nevertheless, all coefficients remain positive, suggesting that the model-level FI ranking is broadly preserved across batch-size configurations.

These findings indicate that the model-level FI estimates are generally robust to batch size, provided that the batch contains more than a single sample. The near-identical rankings obtained for batch sizes of 8 and above suggest that the main interpretative conclusions are not driven by a specific batching choice. At the same time, the reduced agreement observed for batch size 1 confirms that extremely small batches introduce greater variability into the FI estimates. Overall, the results support the reliability of the proposed FI analysis under typical evaluation settings while also highlighting the limitations of singleton-batch explanations.

\subsection{Feature importance vs KAN activation shape}

As our final analysis, we examine the role of the activation functions learned by KAN. As discussed above, feature importance in KAN is driven by the variability of activations. It is therefore meaningful to analyze how the shapes of the learned activation functions influence the estimation of feature importance. Because the proposed KAN module adopts a single-hidden-layer architecture, these activation functions can be visualized directly.

\begin{figure}
    \centering
    \includegraphics[width=1\linewidth]{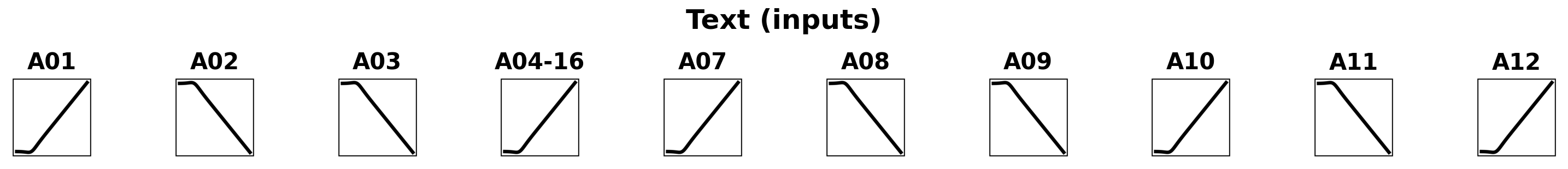}
    \caption{Activation functions connected from the input attribute value \emph{text (inputs)}.}
    \label{fig:KAN_activation_shape_input}
\end{figure}

Based on the global importance ranking of attribute values in Section~\ref{section:exp_res_global_local_fi}, which showed that \emph{text (inputs)} is the most important feature across attacks, we first visualize the activation functions associated with this attribute. Figure~\ref{fig:KAN_activation_shape_input} presents the 10 activation functions corresponding to the 10 attacks involving \emph{text (inputs)}. Overall, these functions exhibit highly similar shapes, with the main differences appearing in their orientation. This consistency suggests that \emph{text (inputs)} contributes in a stable and systematic manner across different attack categories. To be more precise, most of the activation functions follow an approximately monotonic piecewise-linear pattern, indicating that the influence of \emph{text (inputs)} changes in a structured rather than irregular way. The recurrence of nearly identical shapes across attacks further supports the interpretation that KAN captures a common response pattern for this attribute. In other words, although the direction of the effect may vary between attacks, the functional form remains largely preserved. This observation aligns with the high global importance of \emph{text (inputs)}, since a consistently varying activation pattern across tasks implies that this feature provides informative and reusable signals for distinguishing attack types.

\begin{figure}
    \centering
    \includegraphics[width=0.75\linewidth]{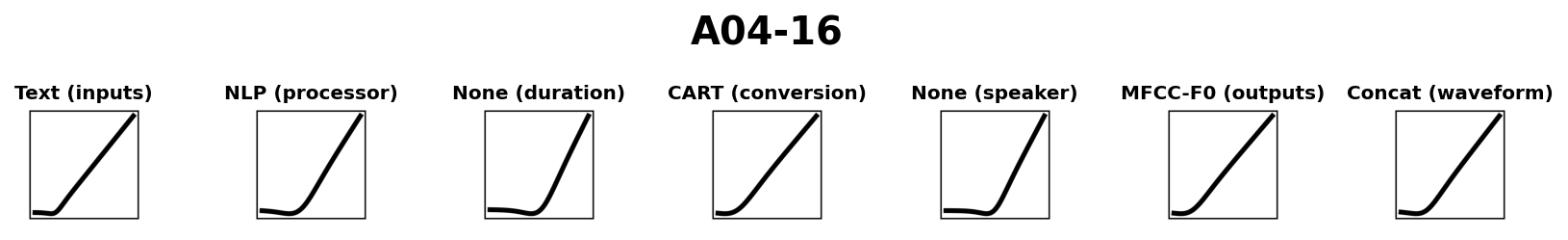}
    \caption{Activation functions connected to the output attack $A04(16)$.}
    \label{fig:KAN_activation_shape_output}
\end{figure}

Next, we visualize the activation functions associated with the attribute values of attack \(A04(16)\), which we select because it shows high consistency and stability in the feature-importance analysis (Figures~\ref{fig:exp_res_FIeval_consistency_diffFImethods_perClass} and~\ref{fig:FI_stability_different_BS_per_attack}). Figure~\ref{fig:KAN_activation_shape_output} graphs the corresponding activation shapes. When compared with the global importance ranking in Table~\ref{tab:global_FI_ranks_per_class_level}, the results suggest an inverse relationship between activation nonlinearity and feature importance: attributes with smoother and more nearly linear activation functions tend to receive higher importance scores. For instance, \emph{text (inputs)}, which exhibits the smoothest and least nonlinear activation pattern, is ranked as the most important attribute for \(A04(16)\).

This finding is consistent with the intuition underlying the KAN feature-importance mechanism, which is based on the standard deviation of the activations. In particular, the activation function for \emph{text (inputs)} shows the largest overall gradient magnitude, which likely produces greater variation in activation values and, consequently, a higher standard deviation than the other attributes. More generally, these results indicate that the shape of the learned activation function is closely related to the importance assigned to the corresponding input feature.

\section{Conclusion}

In this work, we extended our recent probabilistic feature embedding approach for speech deepfake source tracing~\cite{manasi_explainable,mishra_towards}. In contrast to abstract and non-transparent latent-space features, probabilistic attributes provide an interpretable feature domain that reflects the presence or absence of specific synthesizer sub-components. The main novel contributions beyond~\cite{manasi_explainable,mishra_towards} were twofold: (1) an end-to-end training framework, based on multitask learning, for probabilistic attribute extraction; and (2) a structured Kolmogorov--Arnold Network (KAN) for attack classification. Unlike post-hoc explainability methods such as SHAP~\citep{Shapley}, KAN provides a native built-in explainability mechanism. Although our study is not the first study to apply KAN to deepfake source tracing, earlier work in the context of speech deepfakes (Section~\ref{sec:background-XAI}) has primarily used KAN as a replacement for conventional multilayer perceptrons, without fully exploiting the interpretability potential emphasized in the original KAN formulation. In contrast, our proposed design explicitly incorporated this interpretability aspect into the model structure and analysis. Taken together, the proposed model aimed to provide an integrated framework in which classifier decisions and explanations were tightly linked.

Our proposed model achieved balanced accuracies exceeding 99\% on both attribute extraction and attack classification (source tracing) tasks. The corresponding EERs ranged from 0.16\% to 0.07\% for attribute extraction, while attack classification achieved an EER of 0.11\%. 

Beyond classification performance, the proposed framework provided meaningful and robust model explanations. The built-in KAN feature importance mechanism consistently identified interpretable attack attributes and produced rankings that were broadly consistent with those obtained using SHAP. Furthermore, the feature importance estimates remained stable across different evaluation batch sizes, supporting the reliability of the resulting explanations.

Our study addresses the key challenges of moving beyond binary spoofing detection toward fine-grained source tracing, as well as the need for interpretability in speech anti-spoofing. The results demonstrate that spoofed speech attribution combined with a KAN classifier is an effective solution for both accurate classification and transparent decision-making. Moreover, the built-in feature importance mechanism offers a promising alternative to post-hoc explainable AI methods.

Our proposed system has one main limitation that should be addressed in future work. Similar to~\cite{manasi_explainable,mishra_towards}, the attribute extractors rely on available deepfake generator metadata, which limits generalization to scenarios in which such metadata is unavailable. Future work will therefore focus on improving cross-dataset generalization and evaluating robustness under more realistic conditions, including background noise, audio compression, and channel distortions. A further practical consideration is the higher computational complexity of KAN compared to conventional MLP-based architectures. Despite this, our results suggest that KAN provides a promising foundation for developing interpretable and highly accurate speech deepfake source tracing systems. To support transparency, reproducibility, and future extensions, we have made the implementation code publicly available.

\section*{Appendix}

\appendix
\section{Probabilistic composition strategies for RawBoost augmentation} \label{app:data_prep_rawboost}

Table \ref{tab:rawboost_probs_algo} lists the combinations of RawBoost algorithms and their probabilities used during augmentation. Two composition strategies are considered: sequential and parallel composition. The former refers to applying one augmentation to the output of another, whereas the latter generates multiple augmented signals independently from the same input and then combines them. Eight different combinations are considered that involve one, two, or all three main RawBoost algorithms. Since no prior evidence suggests that any particular combination yields greater robustness, we assign all combinations an equal probability of 12\%. The remaining 4\% of the data are kept unchanged to preserve a portion of the original samples.

\begin{table}[htbp]
\centering
\caption{Probabilities of RawBoost augmentation combinations. “$+$” denotes sequential composition, where one augmentation is applied to the output of another, while “$||$” denotes parallel composition, where multiple augmented signals are independently generated from the same input and then combined.}
\label{tab:rawboost_probs_algo}
\renewcommand{\arraystretch}{1.2}
\begin{tabular}{|l|l|c|}
\hline
\multicolumn{1}{|c|}{\textbf{RawBoost algorithms}} & 
\multicolumn{1}{c|}{\textbf{Augmented data}} &
\multicolumn{1}{c|}{\(p_{\text{DA}}\)}  \\
\hline
\(\varnothing\) & \(x\) & 0.04 \\
\hline
LnL & \(\mathrm{LnL}(x)\) & 0.12 \\
\hline
ISD & \(\mathrm{ISD}(x)\) & 0.12 \\
\hline
SSI & \(\mathrm{SSI}(x)\) & 0.12 \\
\hline
LnL + ISD + SSI & \(\mathrm{SSI}(\mathrm{ISD}(\mathrm{LnL}(x)))\) & 0.12 \\
\hline
LnL + ISD & \(\mathrm{ISD}(\mathrm{LnL}(x))\) & 0.12 \\
\hline
LnL + SSI & \(\mathrm{SSI}(\mathrm{LnL}(x))\) & 0.12 \\
\hline
ISD + SSI & \(\mathrm{SSI}(\mathrm{ISD}(x))\) & 0.12 \\
\hline
LnL || ISD & \(\mathrm{LnL}(x) + \mathrm{ISD}(x)\) & 0.12 \\
\hline
\end{tabular}
\end{table}

\section{KAN is configured with auxiliary structure} \label{app:exp_setup_KAN_aux}

\begin{table}[htbp]
\centering
\caption{Connections between input features and hidden-layer activation functions in the SKM. Each connection links an attribute value to its corresponding attack (Figure \ref{fig:asvspoof2019LA_tree_metadata}). The first column shows the input feature index while the second lists the connected function indices.}
\label{tab:skm_connections}
\renewcommand{\arraystretch}{1.0}
\scriptsize	
\begin{minipage}{0.48\textwidth}
\centering
\begin{tabular}{c|l}
\hline
\textbf{Input feature index} & \textbf{Hidden function index} \\
\hline
1  & 1, 2, 3, 4, 7, 8, 9, 10, 11, 12 \\
2  & 5, 6, 16, 17 \\
3  & 13, 14, 15 \\
4  & 1, 2, 3, 4, 7, 8, 9, 12 \\
5  & 5, 13, 16 \\
6  & 6 \\
7  & 10, 11 \\
8  & 14, 15 \\
9  & 17 \\
10  & 1, 2, 8 \\
11 & 3 \\
12 & 7, 9, 12 \\
13 & 10, 11 \\
14 & 13 \\
15 & 4, 5, 6, 14, 15, 16, 17 \\
16 & 1, 2, 8 \\
17 & 3 \\
18 & 4 \\
19 & 5, 16 \\
20 & 6 \\
21 & 7, 9, 12, 14, 15 \\
22 & 10, 11 \\
23 & 13 \\
24 & 17 \\
25 & 1, 2 \\
\hline
\end{tabular}
\end{minipage}
\hfill
\begin{minipage}{0.48\textwidth}
\centering
\begin{tabular}{c|l}
\hline
\textbf{Input feature index} & \textbf{Hidden function index} \\
\hline
26 & 3, 5, 7, 8, 9, 12, 16 \\
27 & 10, 11 \\
28 & 17 \\
29 & 4, 6, 13, 14, 15 \\
30 & 1, 8, 9, 15, 16 \\
31 & 2, 3, 14 \\
32 & 4 \\
33 & 5 \\
34 & 6 \\
35 & 7 \\
36 & 10, 11 \\
37 & 12 \\
38 & 13 \\
39 & 17 \\
40 & 1, 12, 15 \\
41 & 2, 3, 5, 7 \\
42 & 4 \\
43 & 6 \\
44 & 8 \\
45 & 9 \\
46 & 10 \\
47 & 11 \\
48 & 13, 16 \\
49 & 14 \\
50 & 17 \\
\hline
\end{tabular}
\end{minipage}
\end{table}

Table \ref{tab:skm_connections} represents the configuration in the implementation of connections between the input features and the hidden-layer activation functions in the structured KAN module (SKM). The input feature vector is indexed from 0 to 49, with each feature connected to one or more of the 17 possible outputs (indexed from 0 to 16). Each connection maps an input attribute value to the activation functions associated with its corresponding attack, following the hierarchical relationships shown in Figure \ref{fig:asvspoof2019LA_tree_metadata}. In each row, the first column lists the index of the input feature, while the second column specifies the indices of the connected activation functions.

\section{Reproduced baseline source tracing results}\label{app:baseline_results}

The results in Table~\ref{tab:exp_baseline_sourcetracing} summarize the performance of the reproduced two-stage baseline source tracing systems evaluated on the ASVspoof2019-attr-17 protocol. Following the setup of prior studies~\citep{manasi_explainable,mishra_towards}, attack attribute extractors are first trained independently and their output embeddings are subsequently used for attack type classification. We additionally incorporate silence trimming and RawBoost augmentation in all reproduced configurations. The table reports the performance of individual attack attribute extractors ($AS1$--$AS7$) as well as the final attack type classifier using different back-end models, including naive bayes (NB), decision tree (DT), logistic regression (LR), and support vector machine (SVM).

\begin{table}[htbp]
\centering
\caption{Performance of baseline source tracing models (two-phase architectures) on ASVspoof2019-attr-17 protocol. The results on evaluation set are reported as \(<\mathrm{balanced~accuracy}>\) (\%) / \(<\mathrm{equal~error~rate}>\) (\%) . The abbreviations in the table represent the following: CM (countermeasure), BAcc (balanced accuracy), ST (silence trimming), RB (RawBoost), ebds (embeddings), Clf (Classifier).}
\label{tab:exp_baseline_sourcetracing}
\renewcommand{\arraystretch}{2.0}
\scriptsize
\begin{threeparttable}
\begin{tabular}{c|c|c|c|c|c|c|c|c|c}
\Xhline{1.5pt}
CM extractor & 
\multicolumn{7}{c|}{\textbf{Attack attribute classification}} & \multicolumn{2}{c}{\textbf{Attack type classification}} \\
\hline
&\multicolumn{7}{c|}{BAcc (\%) / EER (\%)} & \multirow{2}{*}{Clf} & \multirow{2}{*}{BAcc (\%) / EER (\%)} \\
\cline{2-8}
& $AS1$ & $AS2$ & $AS3$ & $AS4$ & $AS5$ & $AS6$ & $AS7$ & \\
\Xhline{1.5pt}
\multirow{4}{*}{\texttt{ST\_RB\_AASIST}} 
    & \multirow{4}{*}{\makecell[l]{91.59~/\\3.16}}
    & \multirow{4}{*}{\makecell[l]{87.16~/\\3.92}}
    & \multirow{4}{*}{\makecell[l]{85.26~/\\4.58}}
    & \multirow{4}{*}{\makecell[l]{89.51~/\\3.22}}
    & \multirow{4}{*}{\makecell[l]{86.25~/\\5.66}}
    & \multirow{4}{*}{\makecell[l]{83.82~/\\4.00}}
    & \multirow{4}{*}{\makecell[l]{86.06~/\\3.96}}
    & NB
    & 76.70~/~5.41 \\
    \cline{9-10}
& & & & & & & 
    & DT
    & 81.55~/~9.11 \\
    \cline{9-10}
& & & & & & & 
    & LR
    & 84.37~/~3.35 \\
    \cline{9-10}
& & & & & & &  
    & SVM
    & 82.32~/~3.80 \\
    \cline{9-10}
\Xcline{1-1}{1.5pt}\Xcline{2-8}{1.5pt}\Xcline{9-10}{1.5pt}
\multirow{4}{*}{\texttt{ST\_RB\_SSL-AASIST}} 
    & \multirow{4}{*}{\makecell[l]{79.36~/\\10.42}}
    & \multirow{4}{*}{\makecell[l]{70.35~/\\9.75}}
    & \multirow{4}{*}{\makecell[l]{70.16~/\\10.78}}
    & \multirow{4}{*}{\makecell[l]{71.74~/\\9.22}}
    & \multirow{4}{*}{\makecell[l]{67.72~/\\12.99}}
    & \multirow{4}{*}{\makecell[l]{67.09~/\\10.03}}
    & \multirow{4}{*}{\makecell[l]{67.90~/\\9.88}}
    & NB
    & 52.42~/~29.10 \\
    \cline{9-10}
& & & & & & & 
    & DT
    & 58.24~/~12.30 \\
    \cline{9-10}
& & & & & & & 
    & LR
    & 66.23~/~8.48 \\
    \cline{9-10}
& & & & & & &
    & SVM
    & 63.67~/~9.18 \\
\Xhline{1.5pt}
\end{tabular}
\end{threeparttable}
\end{table}

\printcredits

\section*{Declaration of competing interest}
The authors declare that they have no known competing financial interests or personal relationships that could have appeared to influence the work reported in this paper.

\section*{Acknowledgments}
The work has been partially supported by the Academy of Finland (Decision No. 349605, project ``SPEECHFAKES''). The authors wish to acknowledge CSC—IT Center for Science, Finland, for computational resources.

\bibliographystyle{cas-model2-names}

\bibliography{document}






\end{document}